\documentclass[fleqn,usenatbib,useAMS]{mnras}

\usepackage{amsmath}
\usepackage{graphicx}
\usepackage{graphicx,multicol}
\usepackage{times}

\usepackage{amsmath} 

\title[ Mode changing, subpulse drifting and nulling in J2321+6024]{Mode changing, subpulse drifting and nulling in four component conal pulsar PSR J2321+6024 }

\author[Rahaman et al.]{
Sk. Minhajur Rahaman$^{1}$\thanks{E-mail: rahaman@ncra.tifr.res.in}, Rahul Basu$^{2}$, Dipanjan Mitra$^{1,3}$ , George I. Melikidze$^{3,4}$\\
$^{1}$ National Centre for Radio Astrophysics, Tata Institute of Fundamental Research, Post Bag 3, Ganeshkind,Pune-411007,India\\
$^{2}$ Inter-university centre for Astronomy and Astrophysics, Pune-411007, India\\
$^{3}$ Janusz Gil Insitute of Astronomy, University of Zielona G\'ora, ul Szafrana 2, 65-516 Zielana G\'ora, Poland \\
$^{4}$ Evgeni Kharadze Georgian National Astrophysical Observatory, 0301, Abastumani, Georgia }

\begin{document}
\pagerange{\pageref{firstpage}--\pageref{lastpage}}
\maketitle

\begin{abstract}
In this study, we report on a detailed single pulse polarimetric analysis of 
the radio emission from the pulsar J2321+6024 (B2319+60) observed with the 
Giant Metrewave Radio Telescope, over wide frequencies ranging between 300 to 
500 MHz and widely separated observing sessions. The pulsar profile shows the presence of four distinct conal components and belongs to a small group of pulsars classified as a conal quadrupole profile type. The single pulse sequence reveals the presence of three distinct emission 
modes, A, B, and ABN showing subpulse drifting. Besides, there were 
sequences when the pulsar did not show any drifting behaviour suggesting the possibility of a new emission state, which we have termed as mode C. The evolution of the mode changing behavior was
seen during the different observing sessions with different abundance as well as the average duration of the modes seen on each date. The drifting 
periodicities were 7.8$\pm$0.3 $P$, 4.3$\pm$0.4 $P$, and 3.1$\pm$0.2 $P$ in the modes A, B and 
ABN respectively, and showed large phase variations within the mode profile. 
The pulsar also showed the presence of orthogonal polarization modes, 
particularly in the leading and trailing components, which has different characteristics for the stronger and weaker pulses. However, no correlation was
found between the emission modes and their polarization behavior, with the estimated emission heights remaining roughly constant throughout. We have used the Partially Screened Gap model to understand the connection between drifting,
mode changing, and nulling.
\end{abstract}


\section{Introduction}
The coherent radio emission from pulsars is received as periodic single pulses within a 
pulse window. The pulse to pulse variation of single pulses is complex and exhibits a variety of distinct phenomena. A unique signature of any pulsar is the average profile obtained by averaging a few thousand single pulses. The average profile is representative of the emission zone and appears to remain stable across diverse
observation times. However, in a few ($\sim 30$) normal period radio pulsars ($P >$ 0.1 sec), more than one stable emission state exists, 
each of them characterized by distinct emission patterns and average profiles. The phenomenon of transitions between these stable states/modes is 
referred to as mode changing \citep[see e.g.][]{1970Natur.228.1297B,
1971MNRAS.153P..27L,1982ApJ...258..776B,1986ApJ...301..901R}. In general, every
single pulse is structured and consists of one or more sub-components referred to as subpulses. During subpulse drifting, these subpulses exhibit periodic 
variations, either in their position across the pulse window or intensity 
\citep[see e.g.][]{1968Natur.220..231D,1973ApJ...182..245B,1975ApJ...197..481B,
2006A&A...445..243W,2007A&A...469..607W,2016ApJ...833...29B,
2018MNRAS.475.5098B,2019MNRAS.482.3757B}. Also in certain cases nulling 
is seen, where the radio emission goes below the detection
threshold \citep[see e.g.][]{1970Natur.228...42B,1974A&A....31..409H,
1976MNRAS.176..249R,1992ApJ...394..574B,1995MNRAS.274..785V,
2007MNRAS.377.1383W,2012MNRAS.424.1197G,2017ApJ...846..109B}. The transitions to null states are very rapid, usually within one rotation period, but nulling durations vary, ranging from few periods to hours at a time.

\begin{table}
    \centering
     \caption{Parameters of the pulsar PSR J2321+6024 (B2319+60) (taken from \citealt{2005AJ....129.1993M} and \citealt{2004MNRAS.353.1311H}).}
    \begin{tabular}{ccccc}
    \hline 
         Period $P$     & $\dot{\mathrm{P}}$       &    DM   &      RM       & $\dot{\mathrm{E}}$      \\ 
         (s)   &                          & (pc/cc) & (rad/m$^{2}$) & (ergs/s)                \\ \hline
         2.256 &  7.4 $\times$ 10$^{-15}$ &   94.59 &  -232.60      & 2.42 $\times$ 10$^{31}$\\
         \hline 
    \end{tabular}
    \label{tab:pulsar_parameters}
\end{table}

\begin{table*}
\centering
\caption{Observation details for the three runs, 
showing the mode of observation, the time resolution (T$_\mathrm{res}$), the number of pulses recorded (N$_{p}$) and the 50$\%$ and the 10$\%$ widths of the band-averaged integrated profile.   
}
\begin{tabular}{ccccccccc}
\hline
 Date & Obs. Freq & Obs. Mode  & T$_\mathrm{res}$ & N$_{p}$  & $\%$ L & $\%$V & W$_\mathrm{50}$ & W$_\mathrm{10}$ \\ 
      &  (MHz)          &            &  (ms)            &          &       &        &  ($\degr$) &   ($\degr$)      \\ 
\hline
04 Nov, 2017 &  306--339 & Total Intensity & 0.983 & 2122 &    ---   & ---         & 23.5$\pm$0.2 & 28.6$\pm$0.2 \\
16 Nov, 2018 &  300--500 & Full Stokes     & 0.983 & 2489 & 42$\pm$1 & 1$\pm$0.3   & 22.9$\pm$0.2 & 27.8$\pm$0.2 \\
22 Jul, 2019 &  300--500 & Full Stokes     & 0.983 & 2293 & 37$\pm$1 & 1.5$\pm$0.3 & 22.7$\pm$0.2 & 27.6$\pm$0.2\\
\hline  
\end{tabular}
\label{tab:obs}
\end{table*}

The presence of mode changing, subpulse drifting, and nulling is indicative of
the dynamics of the physical processes within the pulsar magnetosphere. The subpulse
drifting is related to the variable $\mathbf{E}\times\mathbf{B}$ drift of the sparking discharges in an inner acceleration region (hereafter IAR) above the polar cap. The non-stationary discharges in IAR produce an outflow of ultra-relativistic charged high energy beams and pair plasma clouds along the open magnetic field lines of the pulsar ( \citealt{1975ApJ...196...51R}; hereafter RS75). The coherent radio emission is attributed to plasma instabilities in the pair plasma clouds. The subpulse represents the average of the emission behavior of thousands of these plasma clouds. In the presence of strong electric and magnetic fields in the IAR, the sparks slightly lag behind the co-rotational motion of the pulsar. This gradual lag is imprinted on the subsequent streams of plasma clouds \citep{2020MNRAS.492.2468M,2020MNRAS.496..465B}. The radio emission mechanism preserves the information about this lag which is seen in the form of subpulse drifting. In contrast to subpulse drifting which is a continuous process, the phenomena of mode changing and nulling represent sudden changes in physical condition in the IAR which alters the sparking process and thereby changes the
radio emission properties. This is further highlighted by recent studies 
\citep{2013Sci...339..436H,2018MNRAS.480.3655H} where the non-thermal component
of the X-ray emission shows correlated changes with the mode changes in the radio emission. Thus, subpulse drifting can be used as a probe to explore the variation of physical condition in the IAR during mode changing. This gives a particularly good motivation to study pulsars which show the presence of subpulse drifting in the different emission modes. 

\begin{figure*}
\begin{tabular}{@{}cr@{}cr@{}cr@{}}
{\mbox{\includegraphics[scale=0.42,angle=0.]{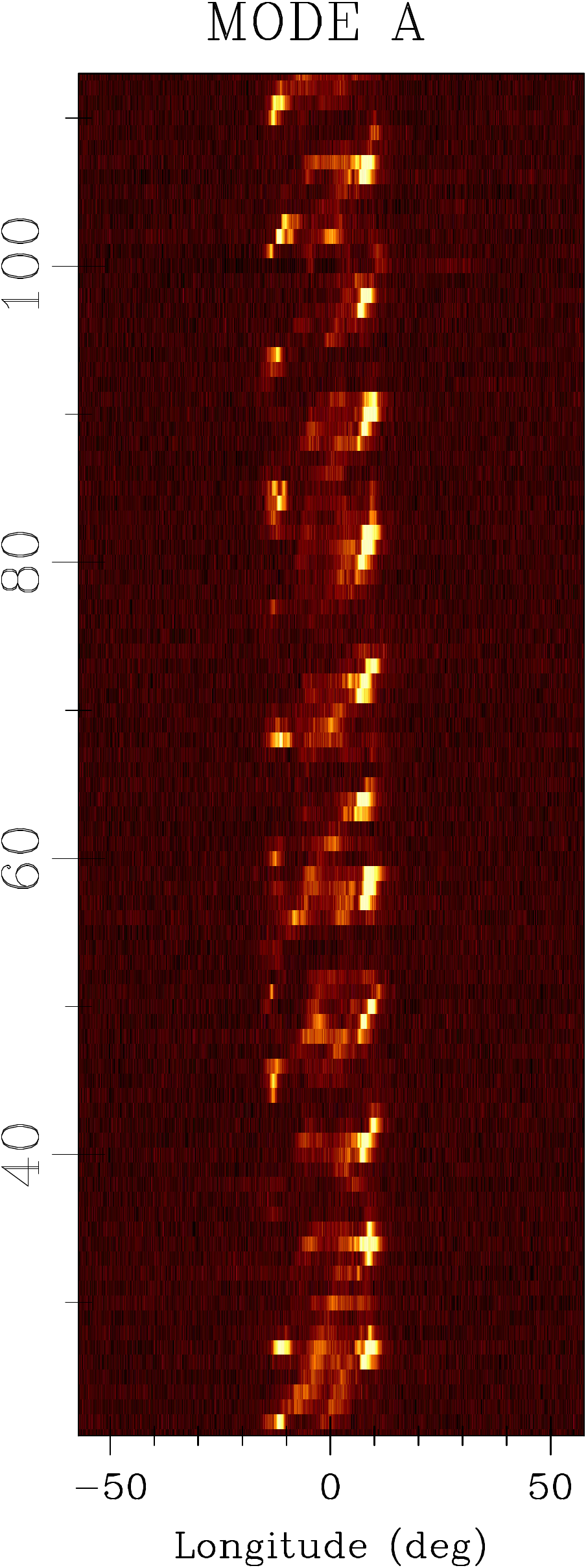}}} &
{\mbox{\includegraphics[scale=0.42,angle=0.]{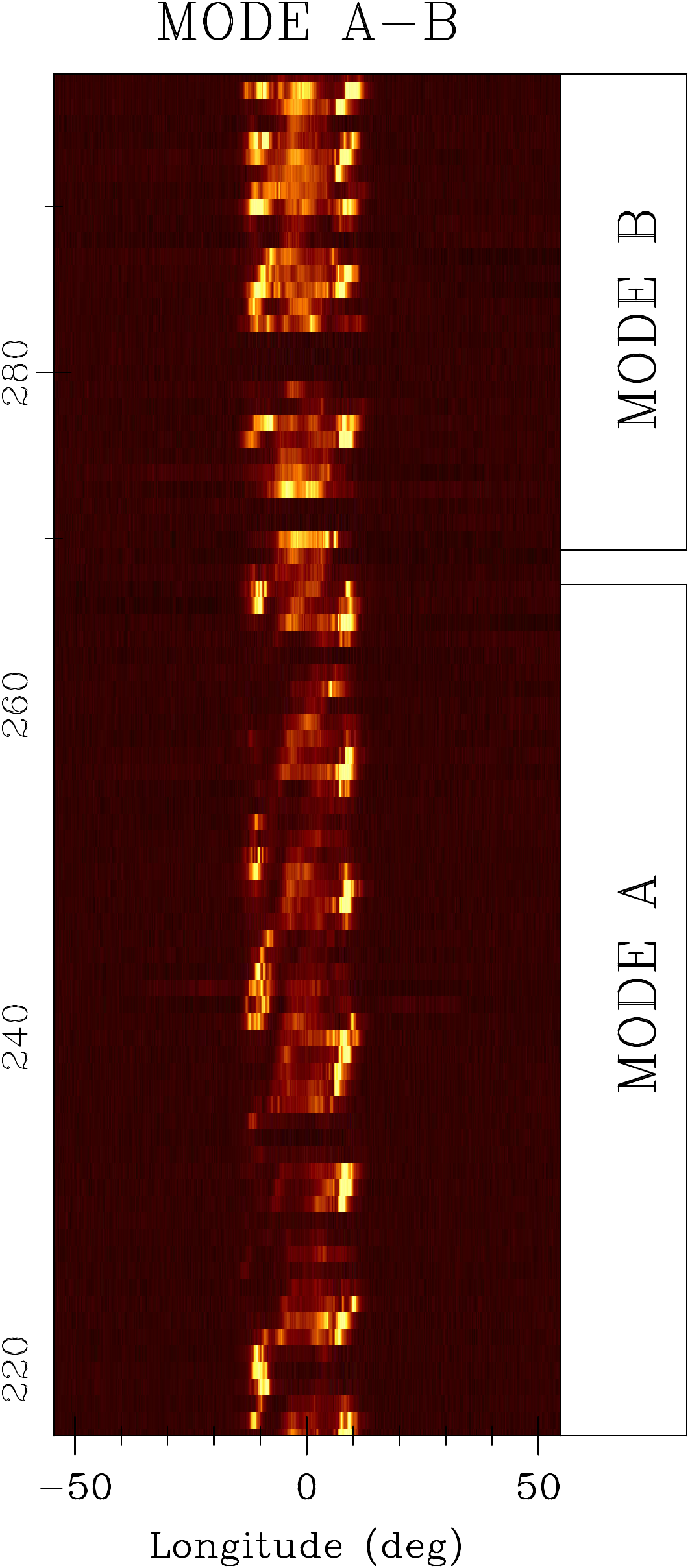}}} &
\hspace{10px}
{\mbox{\includegraphics[scale=0.42,angle=0.]{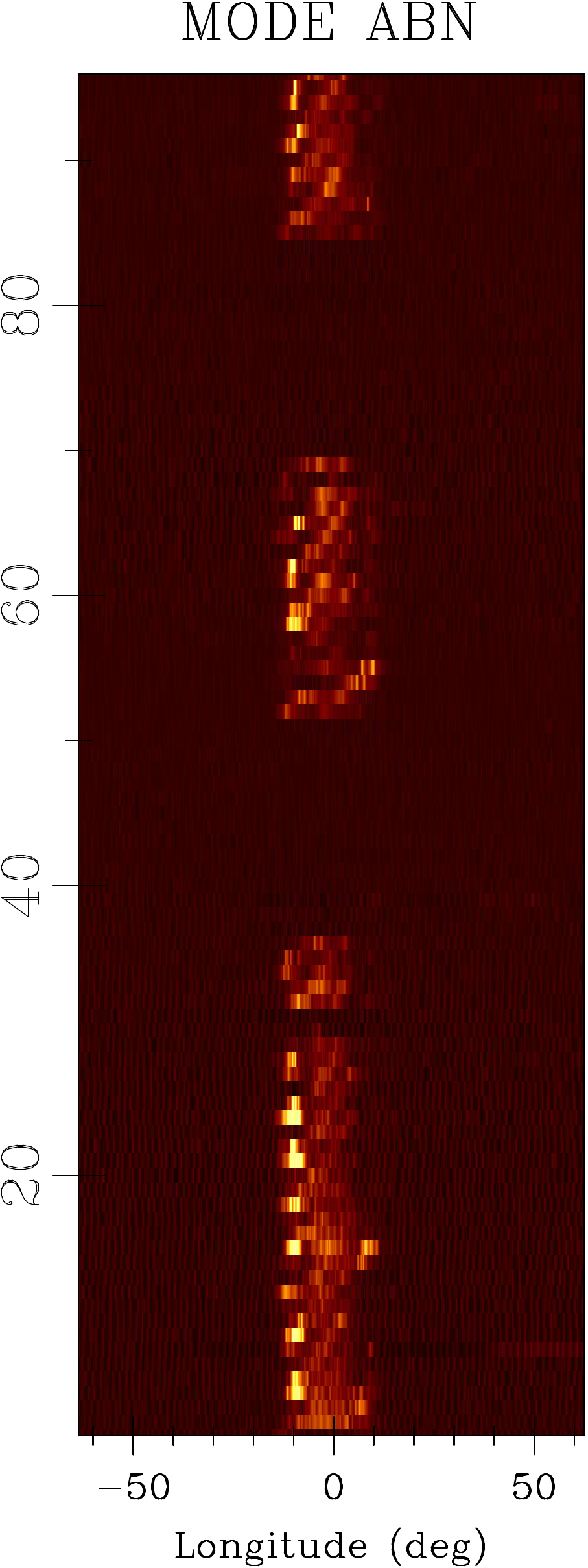}}} &
{\mbox{\includegraphics[scale=0.42,angle=0.]{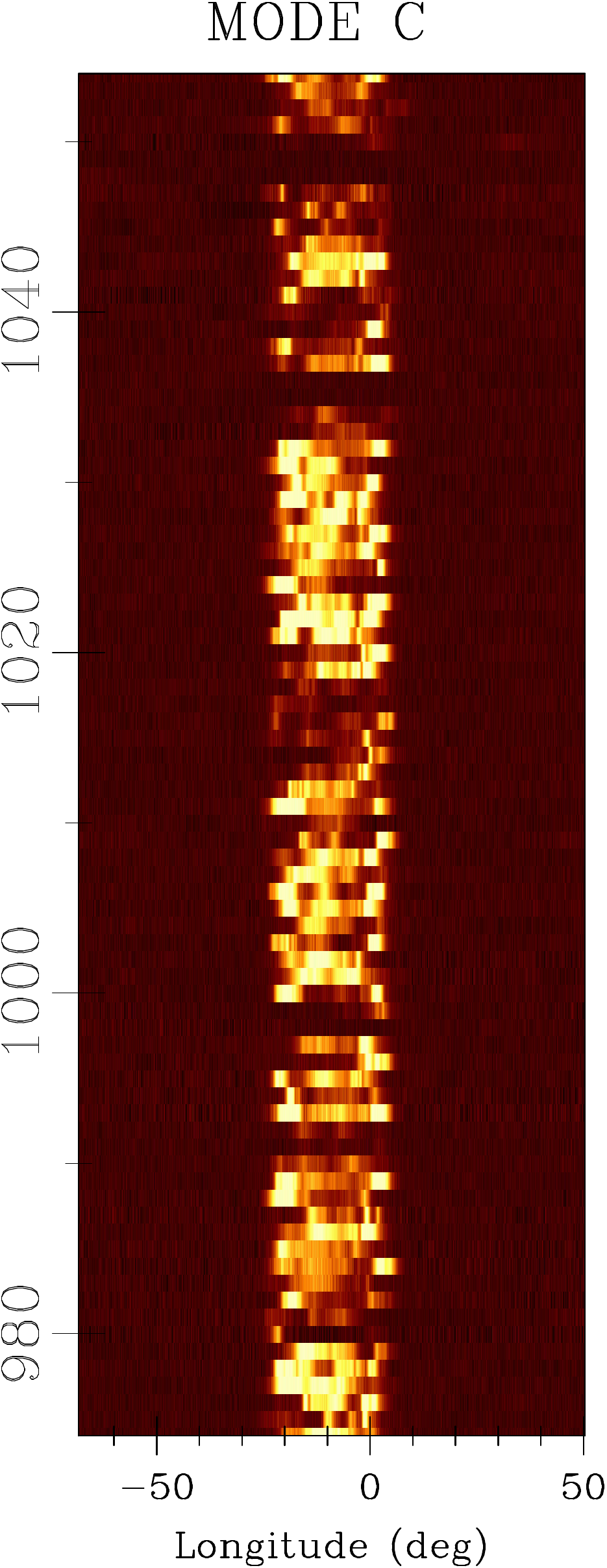}}} \\
\end{tabular}
\caption{The figure shows the four different emission modes seen in the 
single pulse sequence of PSR J2321+6024. The leftmost panel shows a sequence of
mode A on 4 November, 2017 with regular drift bands. The centre-left panel 
shows the an example of transition between modes A and B during 16 November, 
2018. The center-right panel shows single pulses from 22 July, 2019 which shows
the presence of ABN emission mode, interspersed with longer nulls with 
durations $\sim10$-$20P$. The rightmost panel shows a separate pulse sequence 
also from 22 July 2019, which do not show any clear drifting pattern and has 
been identified as a separate mode C.}
\label{fig_modesngl}
\end{figure*}

Only a few pulsars ($<$ 10) show the presence of subpulse drifting in more than one emission mode. Examples include B0031-07 (\citealt{1997ApJ...477..431V} ; \citealt{2005A&A...440..683S}; \citealt{2017ApJ...836..224M}) , B1918+19 (\citealt{1987ApJ...318..410H} ; \citealt{2013MNRAS.433..445R}) , B1944+17 (\citealt{2010MNRAS.408...40K}) , B1237+25 (\citealt{1970Natur.228.1297B} ; \citealt{2005A&A...440..683S} ; \citealt{2013MNRAS.435.1984S} ; \citealt{2014ApJ...792..130M}) , B1737+13 (\citealt{2010MNRAS.406..237F}) , J1822-2256 (\citealt{2018MNRAS.476.1345B}) ,  J2006-0807 (\citealt{2019MNRAS.486.5216B}). Most of these exhibit unresolved conal profiles where the line of sight (LOS) traverses the emission beam towards the edge of the profile and very little evolution is seen in the drifting behavior across the pulsar profile.There are only two known cases where the LOS has central cuts resulting in 
multiple components in the profile, PSR J2006–0807 (five components) and PSR J2321+6024 (four components), and more than two emission modes. The study of the
single pulse behavior of PSR J2006--0807 was carried out by 
\citet{2019MNRAS.486.5216B}. In this work, we have undertaken a comprehensive analysis of the radio emission properties of PSR J2321+6024. The pulsar was 
previously studied as total intensity signals at 1415 MHz by 
\citet{1981A&A...101..356W} (hereafter WF81), where the presence of mode changing, nulling and subpulse drifting in different emission modes were reported. \citet{2014ApJ...797...18G}, studied nulling for this pulsar between 300 MHz and 5 GHz  and established the broadband nature of the phenomenon. We have conducted more sensitive observations at frequencies $<$ 500 MHz, over longer durations at widely separated dates, using
full polarimetric capabilities to study the single pulse behavior in this pulsar. The improved detection sensitivities over a wide frequency bandwidth make it possible to understand the drifting characteristics, 
including the phase evolution, across the entire pulse window. The polarization
properties are intricately related to the emission mechanism and can be used to constrain the radio emission region within the magnetosphere 
\citep{2017JApA...38...52M}. The polarization studies in the different emission
modes are important to explore the effects of mode changes on the emission mechanism. Section \ref{sec:obs} reports the observational details. In section 
\ref{sec:analysis} we present the details of the analysis to characterize the different emission properties of PSR J2321+6024, including details of the emission modes and nulling, subpulse drifting in the different modes, and the properties of the emission region from polarization behaviour as well as average profile. A discussion of the emission properties of PSR J2321+6024 is 
carried out in section \ref{sec:disc} followed by a short summary and 
conclusions in section \ref{sec:sum}.

\section{Observation details} \label{sec:obs}
PSR J2321+6024 was studied using observations from the Giant Metrewave Radio 
Telescope (GMRT), located near Pune in India. The GMRT consists of 30 antennas arranged in a Y-shaped array across 25 km, with 14 antennas in a central square, and the remaining 16 antennas spread on three arms \citep{1991CuSc...60...95S}. 
For high time resolution pulsar studies the interferometer is made to mimic a single dish via 
a specialized observational setup
where signals
from around 20-25 antennas viz., the central square and first few from each arm, are
co-added in the `Phased array' configuration to increase the detection sensitivity of single pulses. The antenna responses are maximized using a 
strong nearby point-like phase calibrator, which is observed at the start of the cycle, and repeated every 1-2 hours. The pulsar was observed on three separate occasions between November 2017 and July 2019. The first observation
was carried out on 4 November 2017 as part of a study on subpulse drifting and
nulling behavior in the pulsar population \citep{2019MNRAS.482.3757B,
2020MNRAS.496..465B}. The pulsars were observed at frequencies between 306-339 
MHz where total intensity emission from each source was measured. We characterized the different emission modes in this initial observation of PSR J2321+6024. Also, we have carried out dedicated observations of this pulsar on 16 November 2018 and 22 July 2019 to
further explore the evolution of the single pulse behavior. These latest observations used the upgraded receiver system of GMRT capable of recording signals over wide frequency bandwidths. We observed in the 300-500 MHz frequency
range and recorded fully polarized emission from the pulsar. The observing duration was around 80-90 minutes on all sessions. While no antenna phasing was
required for the archival data, there was one phasing interval of roughly 5 
minutes between each of the later observations. The basic physical parameters of the PSR J2321+6024 and the observing details are summarized in table \ref{tab:pulsar_parameters} and \ref{tab:obs} respectively.  

A series of intermediate analysis was required to convert the observed time sequence of the pulsar radio emission into a series of properly calibrated 
polarized single pulses. This included removing the data corrupted by radio 
frequency interference (RFI) in time series as well as frequency channels, 
followed by averaging over frequency band, correcting for the dispersion spread 
using the known dispersion measure (see table \ref{tab:obs}) and finally 
re-sampling to get a two-dimensional pulse stack with the longitude on the 
x-axis and pulse number on the y-axis. Besides, the polarization signals which 
were recorded in the auto and cross-correlated form were suitably calibrated 
and converted to the customary four stokes parameters (I, Q, U, V) for each spectral 
channel. The data was divided into several smaller sub-bands of 256 channel each and for each sub-band 
a suitable calibration scheme (see \citealt{2005URSI...J03a}) was applied and finally the phase lags
across the band was corrected for the interstellar rotation measure from \citealt{2005AJ....129.1993M} (see table \ref{tab:pulsar_parameters}). 
The post-calibration details of the analysis for both the total intensity and full polarization pulsar 
observations are reported in \citep{2016ApJ...833...28M}. 
The wideband
observations were further subdivided into five separate parts centered around 
315 MHz, 345 MHz, 395 MHz, 433 MHz, and 471 MHz, to study the frequency evolution of pulsar emission. The five sub-bands had equal bandwidth of 30 MHz but are not equispaced due to the presence of RFI in-between. 

\section{Analysis and Results} \label{sec:analysis}

\subsection{Emission Modes}\label{sec:mode}

\begin{figure*}
\begin{tabular}{@{}cr@{}}
{\mbox{\includegraphics[scale=0.65,angle=0.]{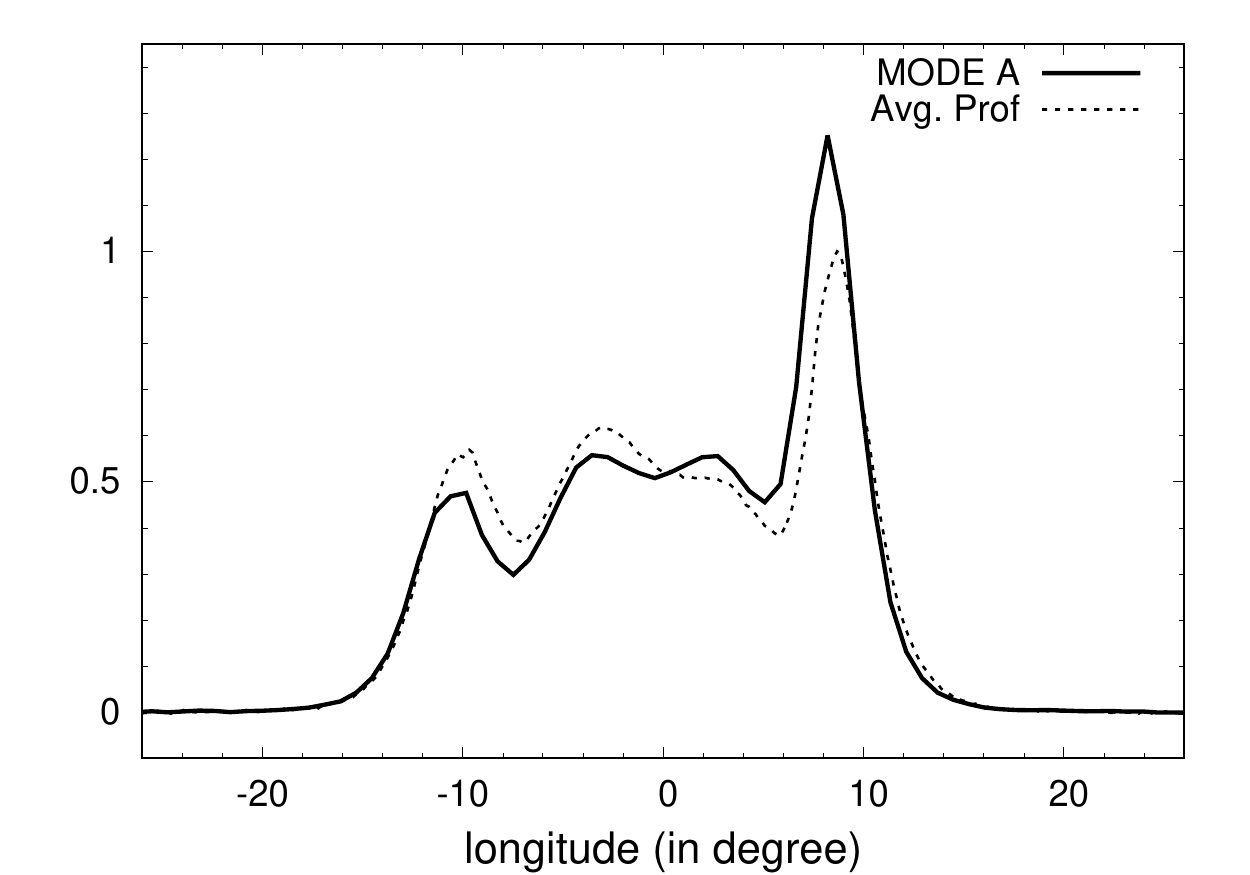}}} &
{\mbox{\includegraphics[scale=0.65,angle=0.]{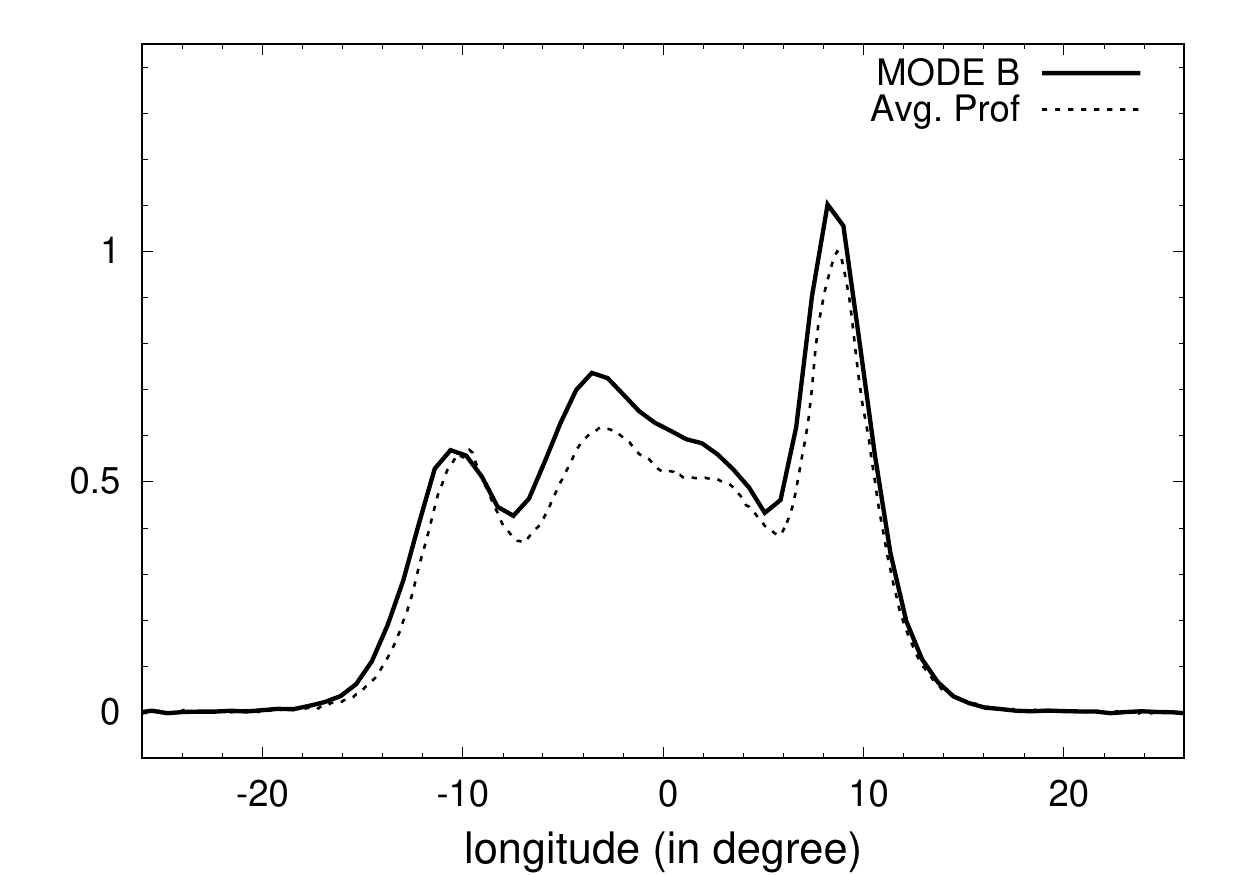}}} \\
{\mbox{\includegraphics[scale=0.65,angle=0.]{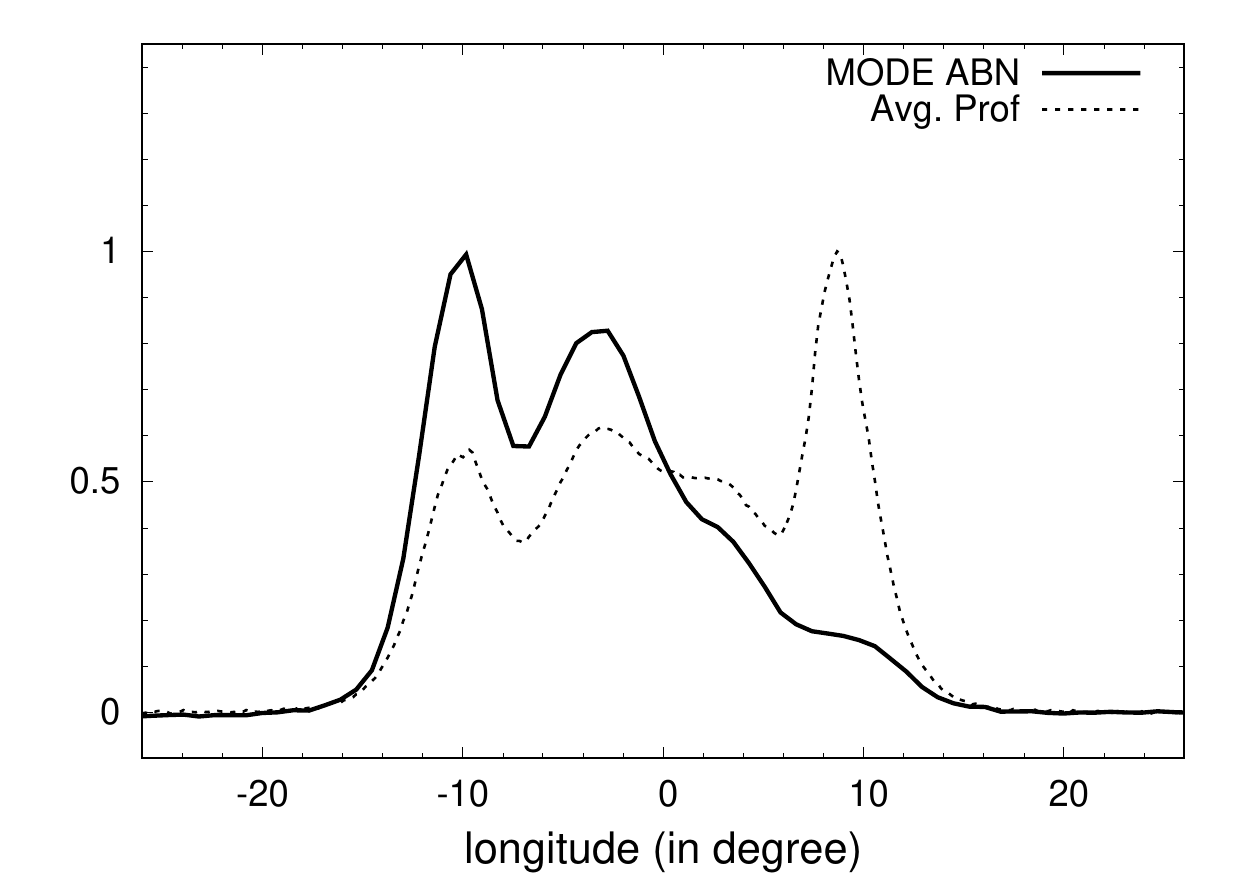}}} &
{\mbox{\includegraphics[scale=0.65,angle=0.]{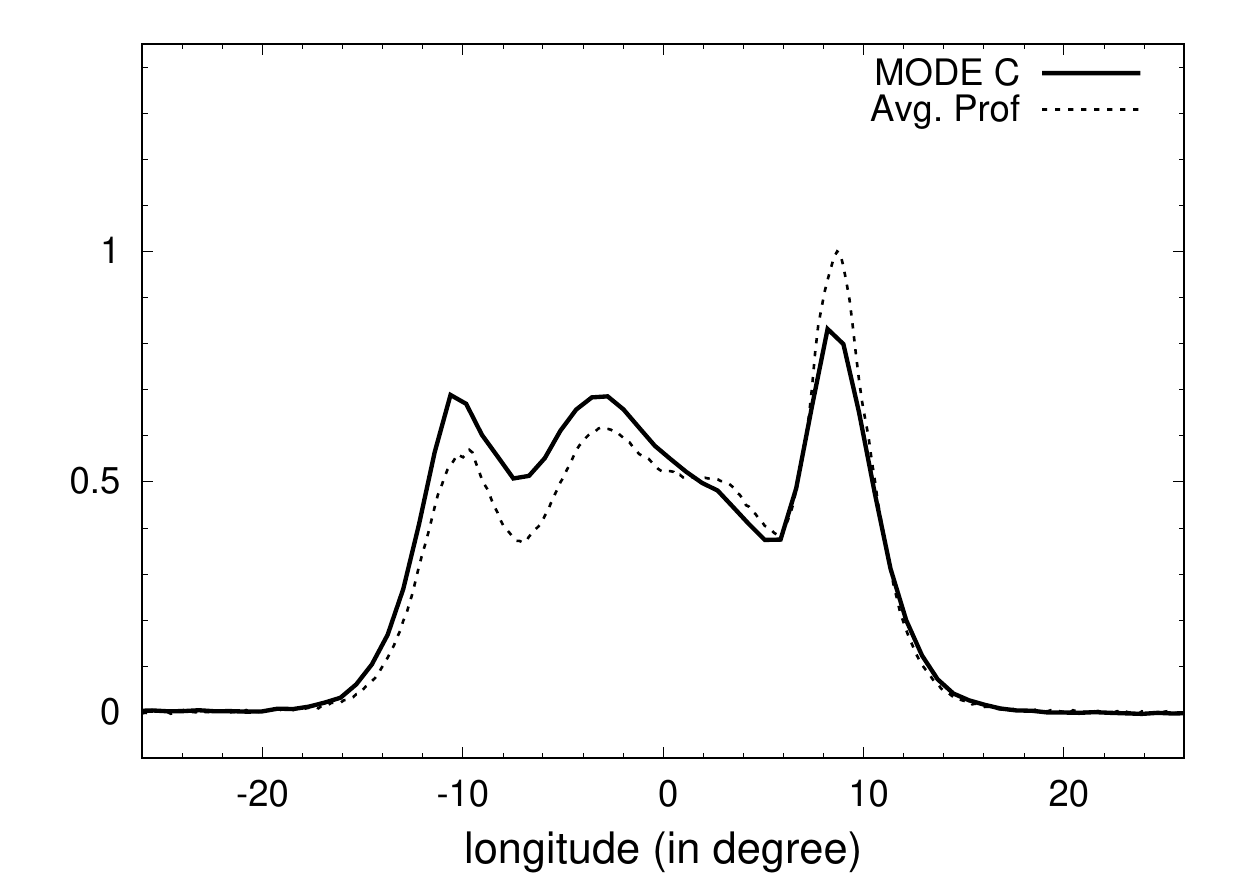}}} \\
\end{tabular}
\caption{The figure shows the average profile of the three emission modes in 
PSR J2321+6024, A (top left panel), B (top right panel), ABN (bottom left 
panel) and C (bottom right panel). The average profile is also shown in each 
plot where the intensities are normalized by the peak value of the average 
profile. The measurements correspond to the observations on 16 November, 2018 
and the frequency sub-band centred around 395 MHz.} 
\label{fig_modeprof}
\end{figure*}

The single pulses in PSR J2321+6024 show the presence of multiple emission modes with distinct physical characteristics. The high detection sensitivity of
single pulse emission on all three sessions enabled us to identify the emission
modes by visually inspecting the pulse sequence. The emission was erratic, 
switching between longer duration modes lasting between 50 to 100 $P$, 
interspersed with short bursts of less than 10 $P$. Nulling was also seen prominently in the pulse sequence. WF81 identified three modes which they classified as modes A, B, and ABN. All three modes exhibited subpulse drifting.
We have also found the presence of these emission states and have adopted the above nomenclature. Besides these emission modes, there were certain sequences which did not show 
the presence of subpulse drifting, suggesting the possibility of a new mode.

The most abundant mode was classified as mode A and showed the presence of distinct subpulse drifting with periodicity $\sim8\;P$. The emission was most prominent in the trailing component (4$^{th}$ component) where the intense bursts of emission were bunched up and repeated at regular intervals, (see Fig. \ref{fig_modesngl}, leftmost panel). The leading component (1$^{st}$ 
component) also showed similar bunching and repetition, but was much weaker than the trailing component. A roughly 180\degr~phase difference was seen between the leading and the trailing components with bunched emission alternating in the two parts. The central components showed systematic drift bands with large phase variations across the entire component. The average profile of mode A is shown in Fig. \ref{fig_modeprof}, top-left panel, which shows that the emission is stronger in the trailing part and weaker in the other three components, compared to the average profile. 

The second drifting mode seen in the pulse sequence was classified as mode B. 
This was primarily a short duration mode lasting typically between 10 to 20 
$P$. There were frequent short duration nulls within this mode. The 
emission was bunched up in the leading and trailing components, similar to the mode
A, which repeated every 4-5 $P$. An example of the mode B is shown in Fig. 
\ref{fig_modesngl}, center-left panel, where the transition from mode A to the mode B is seen along with the short nulls. The average profile, in Fig. 
\ref{fig_modeprof}, top-right panel, shows the emission to be stronger in the trailing component but is also higher than the average value in the central parts. 

The third drifting mode, called the ABN mode, was also seen for short durations lasting between 10 to 20 $P$. Fig. \ref{fig_modesngl}, 
center-right panel shows three sequences of mode ABN separated by nulls. The 
emission was primarily seen towards the leading part of the pulse window and 
was negligible towards the trailing edge, as highlighted in the average profile
(Fig. \ref{fig_modeprof}, bottom-left panel). The presence of systematic drift bands was visible with periodicity around 3 $P$. 

There was also the presence of an additional state which did not show any 
clear drifting behavior and was classified as a possible new mode C. The emission in this mode 
was seen at multiple timescales, lasting from around 10 $P$ to 50-100 
$P$ at a time. The mode C  had short duration nulls interspersed with the emission. Fig. \ref{fig_modesngl}, rightmost panel, shows an example of this state, where unlike the cases the emission was seen throughout the pulse window. This is also highlighted in the profile, Fig. \ref{fig_modeprof}, 
the bottom-right panel, where the leading part is comparable to the fourth component. Additionally, the emission in the central components was merged without any 
clear distinction between them. It is possible that mode C is a mixed state mostly comprising of mode B pulses mixed with mode ABN pulses. A detailed breakdown of each 
emission mode present during all three observing sessions is shown in Table \ref{tab:modelist} in   
Appendix, where the pulse range from the start and the duration of the modes 
are shown sequentially.

\begin{table}
\centering
\caption{Mode statistics for the three different observing sessions, 
showing the relative abundance of each mode, the number of mode sequences 
(N$_\mathrm{T}$) and the average mode lengths.}
\begin{tabular}{ccccc}
\hline
 Date & Mode & Abundance & N$_\mathrm{T}$ & Avg. Length \\ 
      &      &    (\%)   &       &    ($P$)    \\ 
\hline
 04 Nov, 2017 &  A  & 44.0 & 12 & 77.8 \\
              &  B  & ~5.9 & 10 & 12.6 \\
              & ABN & ~4.8 & ~7 & 14.6 \\
              &  C  & 18.5 & 16 & 24.5 \\
              &     &      &    &      \\
 16 Nov, 2018 &  A  & 33.6 & 22 & 38.0 \\
              &  B  & 14.7 & 25 & 14.7 \\
              & ABN & ~7.3 & 17 & 10.6 \\
              &  C  & 11.4 & 14 & 20.2 \\
              &     &  &  &  \\  
 22 Jul, 2019 &  A  & 32.6 & 24 & 31.1 \\
              &  B  & 14.0 & 21 & 15.3 \\
              & ABN & ~8.9 & 18 & 11.3 \\
              &  C  & 11.8 & ~9 & 30.0 \\
\hline  
\end{tabular}
\label{tab:modstat}
\end{table}
    
\begin{figure}
\centering
\includegraphics[scale=0.5,angle=0.]{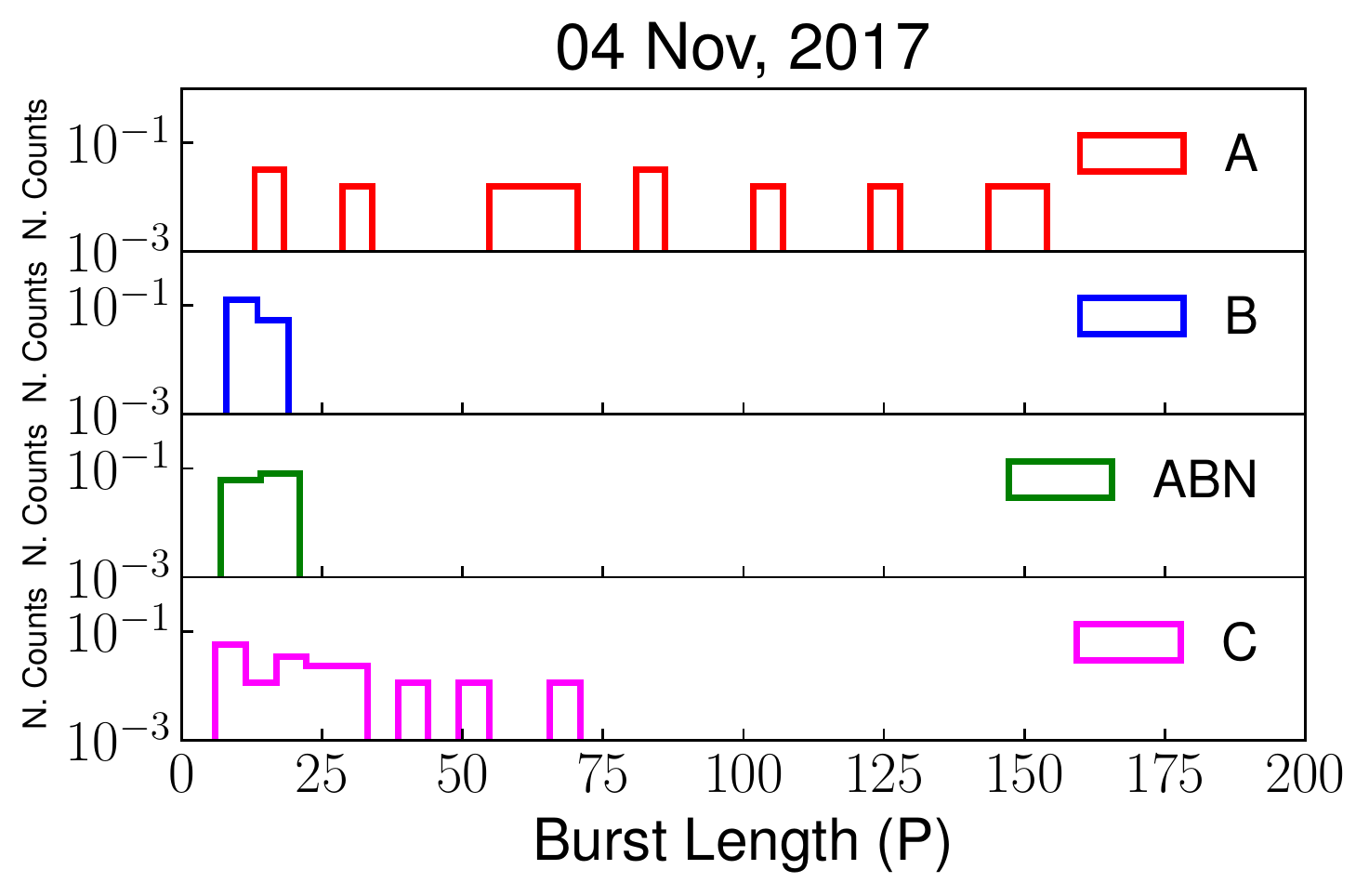}\\
\includegraphics[scale=0.5,angle=0.]{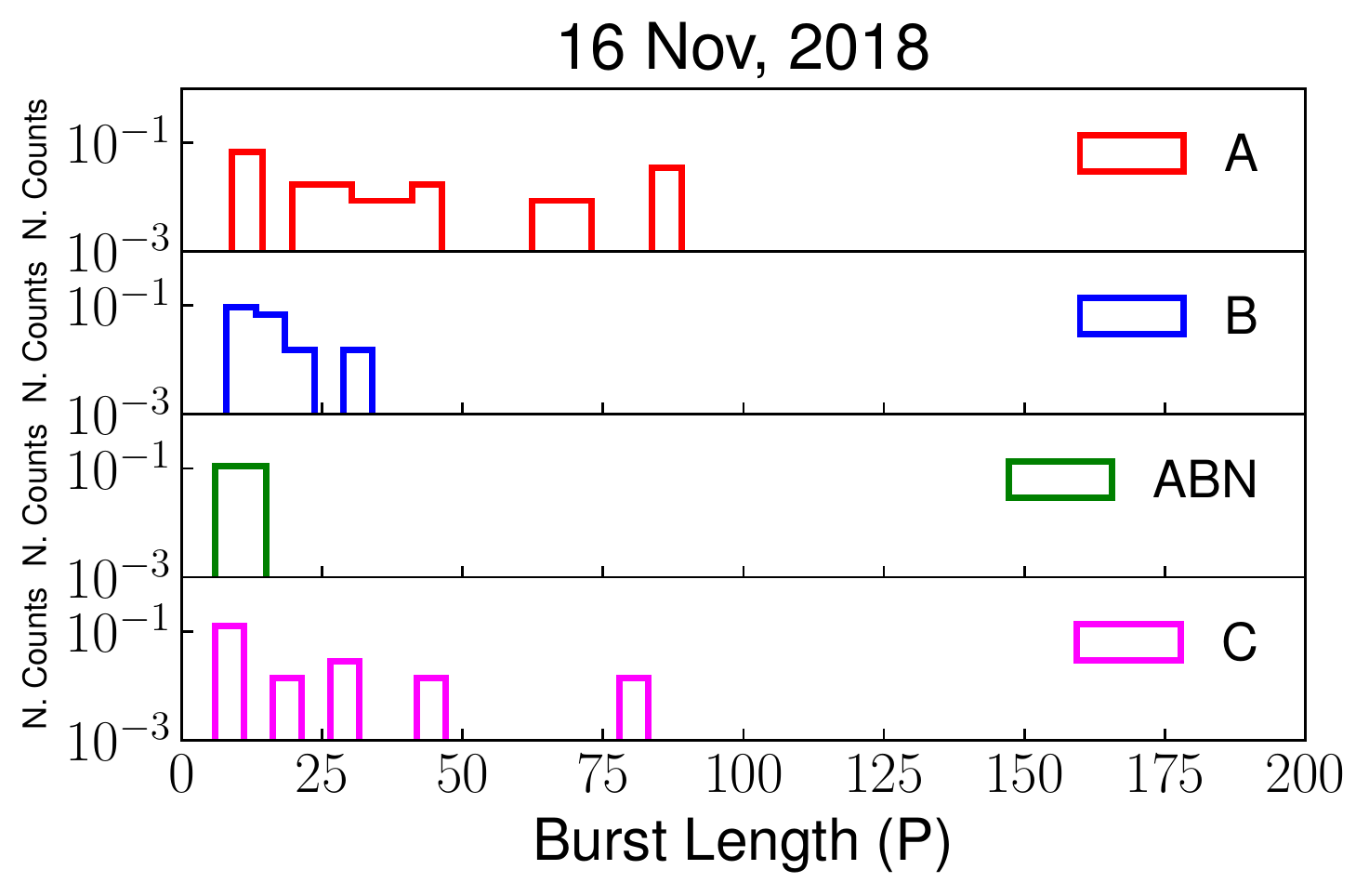} \\
\includegraphics[scale=0.5,angle=0.]{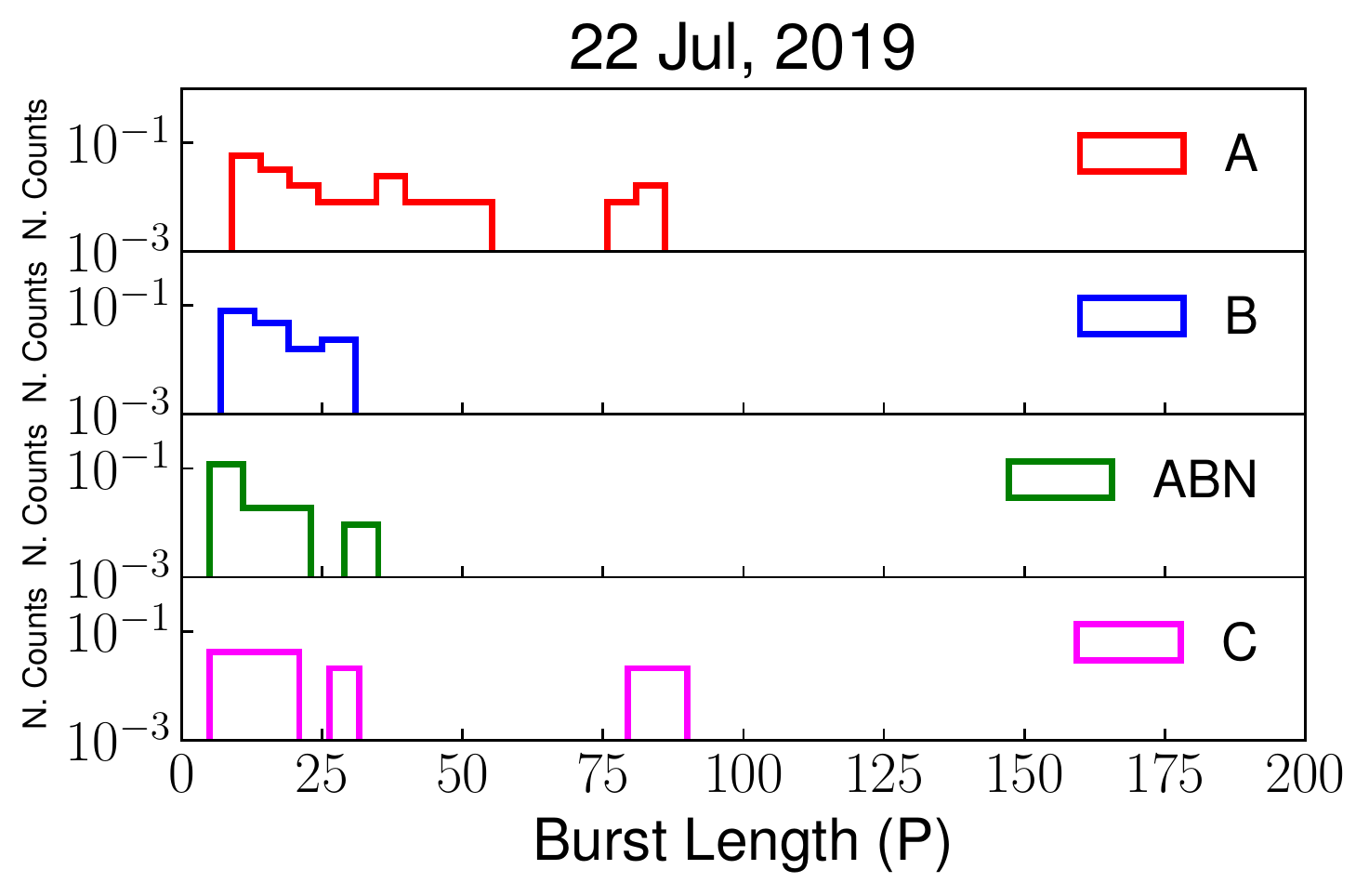} \\
\caption{The figure shows the distributions of the burst length of modes A, B, ABN, and C on three different dates of observations. The y-axis has been normalized by the total number of pulses during each observing session.}
\label{fig_modelen}
\end{figure}

\begin{figure*}
\centering
\begin{tabular}{@{}cr@{}}
{\mbox{\includegraphics[scale=0.7,angle=0.]{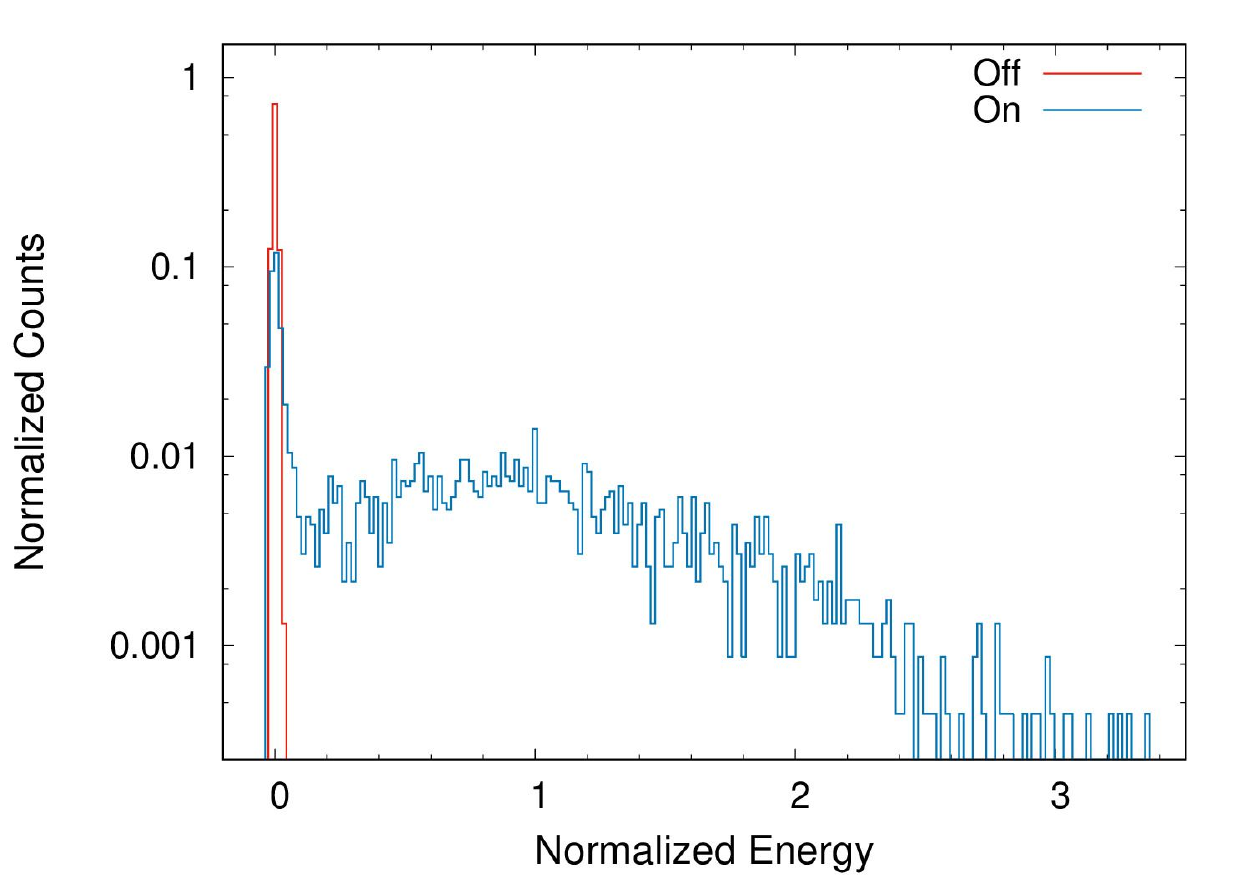}}} &
{\mbox{\includegraphics[scale=0.75,angle=0.]{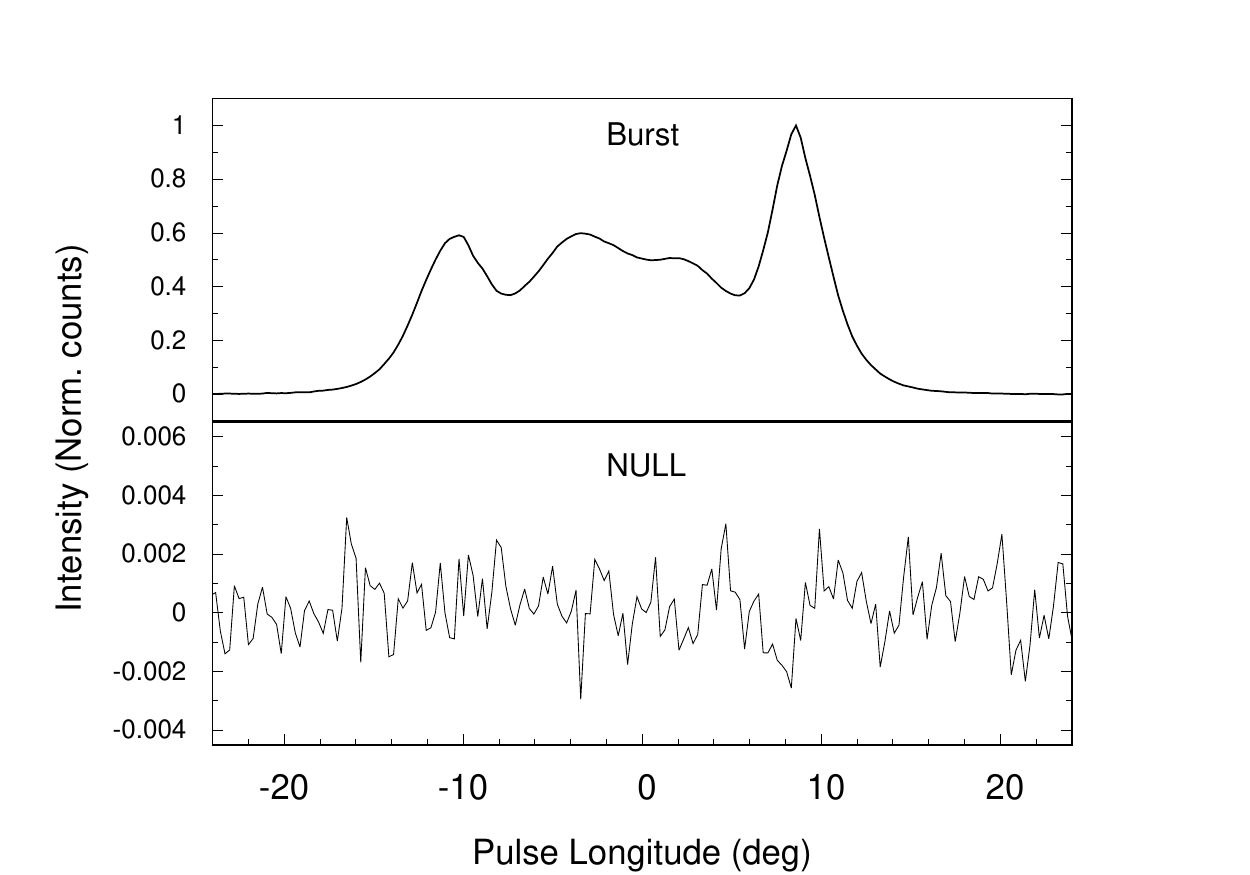}}} \\
\end{tabular}
\caption{The left panel shows the distribution of the energies in the pulse 
window (blue) and the noise levels (red) in an off-pulse region away from the 
pulse window for the observation on 16 November 2018. The pulsar emission
shows a bimodal distribution corresponding to nulling and the burst state, 
where the null distribution is coincident with the off-pulse histogram. The 
x-axis has been normalized by the average energy across all pulses, while the 
y-axis is normalized by the total number of pulses. There is a clear 
statistical boundary between the null and burst states which is highlighted in 
the folded profiles shown in the right panel, which were formed by averaging 
the null (bottom window) and burst (top window) pulses separately. There is no
indication of any low-level emission present during the null states.}
\label{fig_distprof}
\end{figure*}

Table \ref{tab:modstat} shows the average behavior of the emission modes during each observing session, including the percentage abundance, the total number of modes (N$_\mathrm{T}$), and the average duration. The distributions of the duration of modes A, B, ABN, and C are shown in Fig. \ref{fig_modelen}. The table highlights the general tendencies of the four emission states on the
three observing dates, 4 November 2017 
(D1), 16 November 2018 (D2) and 22 July 2019 (D3) of similar observing duration between 2100-2500 $P$. Mode A was the most abundant state with a percentage abundance of 44 \% on D1 with an average duration of 77.8$P$. This reduced to 33.6\% on D2 and 32.6\% on D3, while the average duration decreased by more than 50\% at 38.0$P$ and 31.1$P$ respectively. In contrast mode B was
least abundant on D1 at 5.9\% which roughly tripled to 14.7\% on D2 and 14.0\% 
on D3. However, unlike mode A the average duration did not show much change over the three days, ranging between 12.6$P$, 14.7$P$, and 15.3$P$. Mode ABN 
showed similar evolution to mode B and was least abundant on D1 at 4.8\% 
which increased to 7.3\% on D2 and 8.9\% on D3. The average duration were roughly constant at 14.6$P$, 10.6$P$, and 11.3$P$ on D1, D2, and D3. Finally, 
mode C was most abundant on D1, seen for 18.5\% of the observing duration, 
which reduced to 11.4\% on D2 and 11.8\% on D3. The average mode lengths varied
between 24.5$P$, 20.2$P$, and 30.0$P$, respectively.

To check if the distributions of the emission modes on the 
different observing sessions are similar we have used the Kolmogorov-Smirnov 
(K-S) test, which applies to non-parametric, unbinned distributions 
\citep{1992nrca.book.....P}. This involves estimating the cumulative distribution function of each set of numbers (mode lengths in our case) and determine the parameter $D$, which is defined as the maximum value of the absolute difference between the two cumulative distribution functions. The $D$
factor along with the total number of data points in each distribution is used
to determine the probability ($p$) that the two sets of numbers are part of the identical underlying distribution function.The K-S test was conducted for each emission mode using the set of numbers 
corresponding to the duration of modes on each observing session as reported in
Table \ref{tab:modelist} in the Appendix. For example, on D1 there were 12 different occasions the emission was in mode A with varying durations. Similarly, there were 22 
separate occasions when the emission was in mode A during D2. The 12 numbers on
D1 form one distribution while 22 numbers on D2 form another set of distribution. The K-S test was performed on these two distributions and the
probability of the two being from the same distribution was 3.8$\%$. The probability that the 12 mode A durations on D1 and 24 durations on D3 were part of the same distribution was 0.2$\%$. In contrast, it was much more
likely with 80.2$\%$ probability that the sequences on D2 and D3 were from the same distribution. In the first session mode A was usually seen for longer durations, which were not interrupted by long nulls. In contrast, the durations were shorter during the later observations. On the second day, there were 3 
cases when A to A transitions were reported, while on 7 occasions A to A 
transition was reported on the third day. In each case the two A modes were separated by long nulls ranging between 10 to 40 periods with a lack of continuity in the drifting behaviour between them, prompting the separate classification. For the other modes, there was no evidence for a significant difference in the distributions.

\subsection{Nulling}\label{sec:null}

\begin{table}
\caption{Nulling behaviour during the two wide band observations as 
specified by the nulling fraction (NF), the significance of the nulling 
detection using the parameter $\eta$ which is defined in the text, the number 
of transitions (N$_\mathrm{T}$) between the null and burst states during each 
observing run, and the average duration of the burst ($\langle BL\rangle$) and 
null ($\langle NL\rangle$) states.}
\label{tab:nullstat}
\begin{tabular}{cccccc}
\hline
 Date & NF & $\eta$ & N$_\mathrm{T}$ & $\langle BL\rangle$ & $\langle NL\rangle$ \\
   & ($\%$) &   &   & ($P$) & ($P$) \\ 
\hline 
 16 Nov, 2018 & 35.9$\pm$1.2 & 1.7$\times$10$^4$ & 118 & 13.4 & 7.5 \\
 22 Jul, 2019 & 34.9$\pm$1.2 & 1.0$\times$10$^4$ & 102 & 14.5 & 7.5 \\
\hline                   
\end{tabular}
\end{table}

\begin{figure}
\centering
\includegraphics[scale=0.75,angle=0.]{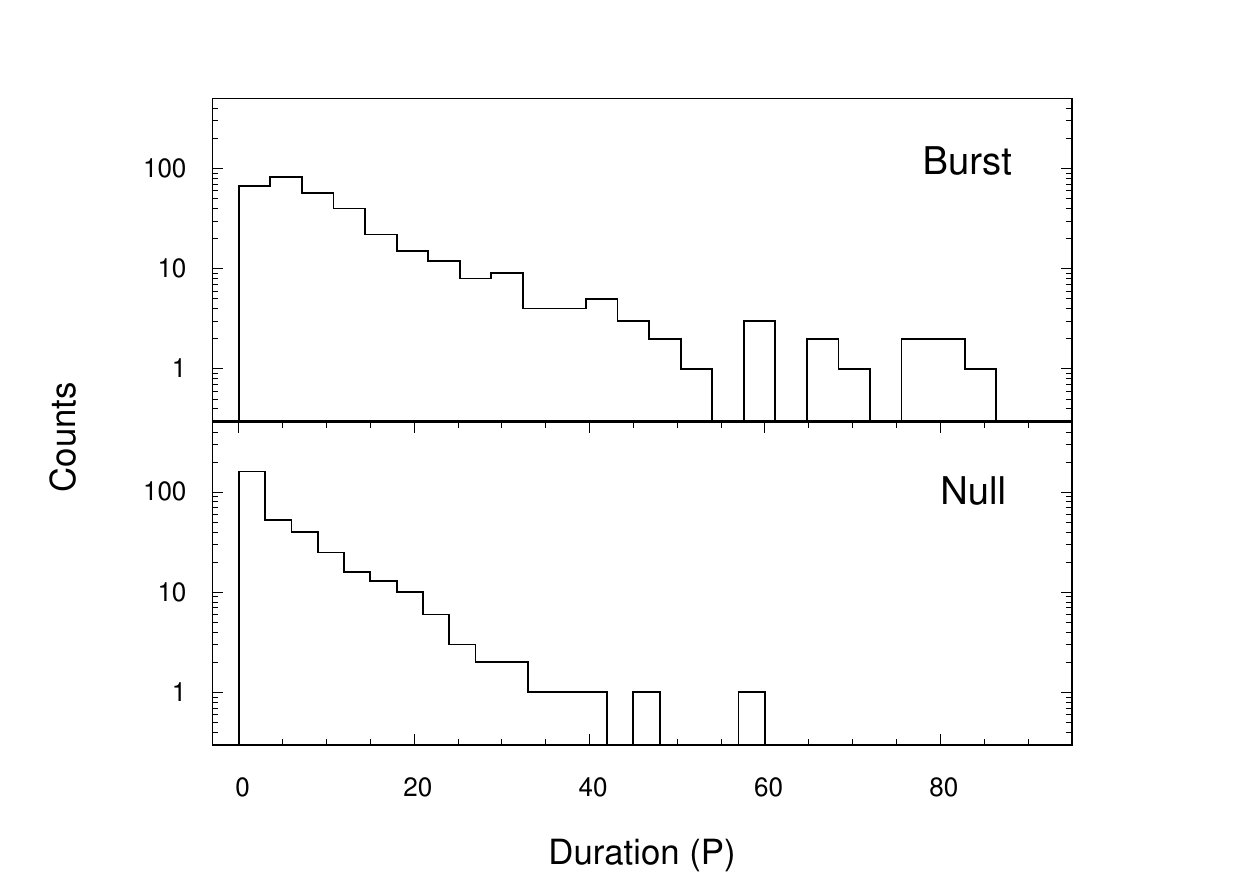} 
\caption{The figure shows the distribution of the durations of the null and
the burst states from all three observing sessions.}
\label{fig_lendist}
\end{figure}

\begin{figure}
\centering
\includegraphics[scale=0.73,angle=0.]{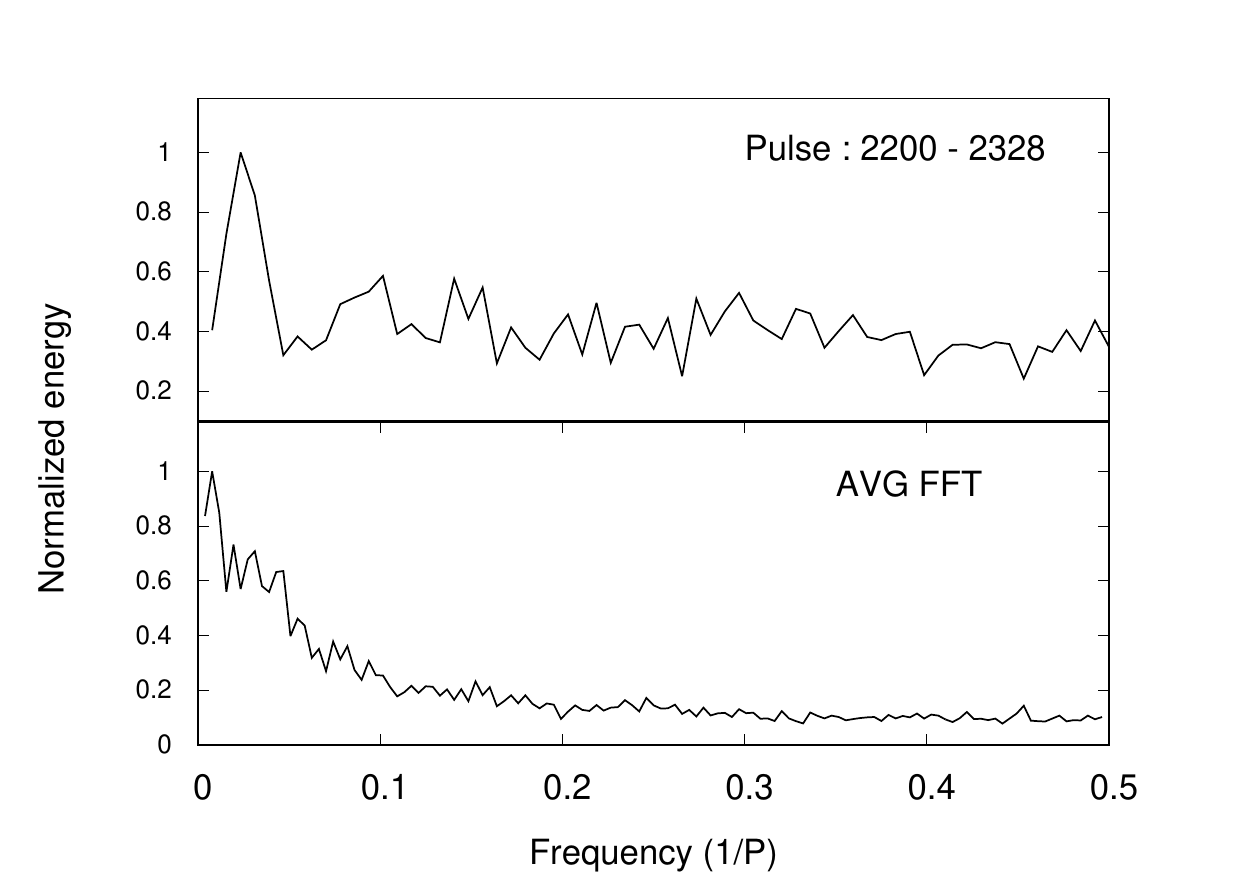} 
\caption{The figure shows the periodicity estimates of nulling for the 
observations on 22 July 2019. A binary time series was setup where the nulls 
were identified as `0' and the bursts as `1'. Fourier transform of the binary 
series was carried out to estimate the periodicity associated with nulling. The
top window shows an example of the Fourier spectra between pulse 2200 and 2328
with prominent periodic behaviour. The lower panel shows the average FFT 
obtained using 256 consecutive numbers at a time and repeating the process for
the entire series by shifting the start position.}
\label{fig_nullfft}
\end{figure}

A study of the nulling behavior for the observations on 4 November 2017 
was reported in \citet{2020MNRAS.496..465B}. We have carried out similar 
estimates for the later two observations. The wideband frequency observations with very high sensitivity of detection of emission, made it possible to statistically separate the null and the burst emission states. The distribution
of the average energies in each single pulse window, shown in Fig. 
\ref{fig_distprof} (left panel) for the 16 November 2018 observations, clearly shows the separation between the null and the burst distributions. The null pulses were identified statistically and the nulling behavior was characterized
in Table \ref{tab:nullstat}. The nulling fraction (NF) was estimated to be around 35\% on both days. The average profile from the null pulses (see Fig. 
\ref{fig_distprof}, right panel) did not show  the presence of any low-level 
emission during nulling. The quantity $\eta$, which is the ratio between the 
total energy within the pulse window during the bursting states (top window in 
Fig. \ref{fig_distprof}, right panel) and the noise level ($3\sigma$) in the null profile (bottom window in Fig. \ref{fig_distprof}, right panel), measures
the extent of radio emission being diminished during nulling. The estimated 
$\eta$ is between 1-2$\times10^4$, which is at least five times higher than the
previously reported values \citep{2012MNRAS.424.1197G,2019MNRAS.486.5216B}, as expected for the wide frequency bandwidths used. The table also reports the 
number of transitions (N$_\mathrm{T}$) from the null to the burst states and 
the average duration of the null phase ($\langle NL\rangle$) and the burst phase ($\langle 
NL\rangle$). Fig. \ref{fig_lendist} shows the distribution of the null length and burst lengths which clearly shows the prevalence of short duration nulls in
addition to the longer null sequences. \cite{2020MNRAS.496..465B} also reported
the nulling to show periodicity during transitions from the null to the burst states. We have identified the null and burst pulses as a binary series of 
`0/1' and estimated the time-evolving Fourier spectra 
\citep{2017ApJ...846..109B}. The nulling shows quasi-periodic behaviour with certain intervals showing more prominent periodic features. Fig. 
\ref{fig_nullfft} top panel shows an example of a pulse sequence on 22 July 
2019, where the nulling periodicity was more prominent. The bottom panel in the figure shows the average spectra over the entire observing duration. The estimated periodicity were between 50-100 $P$.

\subsection{Subpulse Drifting} \label{sec:drift}

\begin{figure*}
\begin{tabular}{@{}cr@{}}
{\mbox{\includegraphics[scale=0.4,angle=0.]{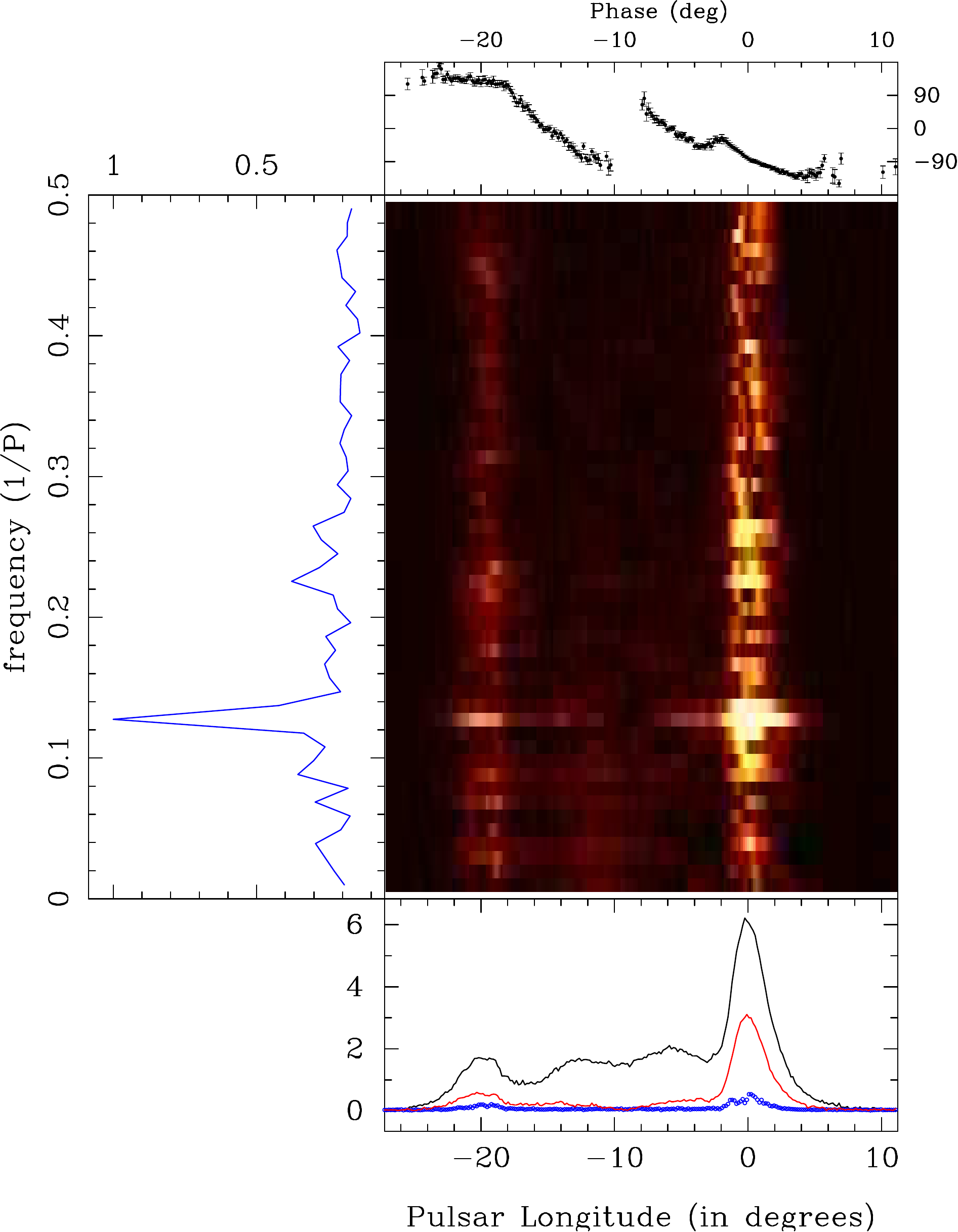}}} &
{\mbox{\includegraphics[scale=0.4,angle=0.]{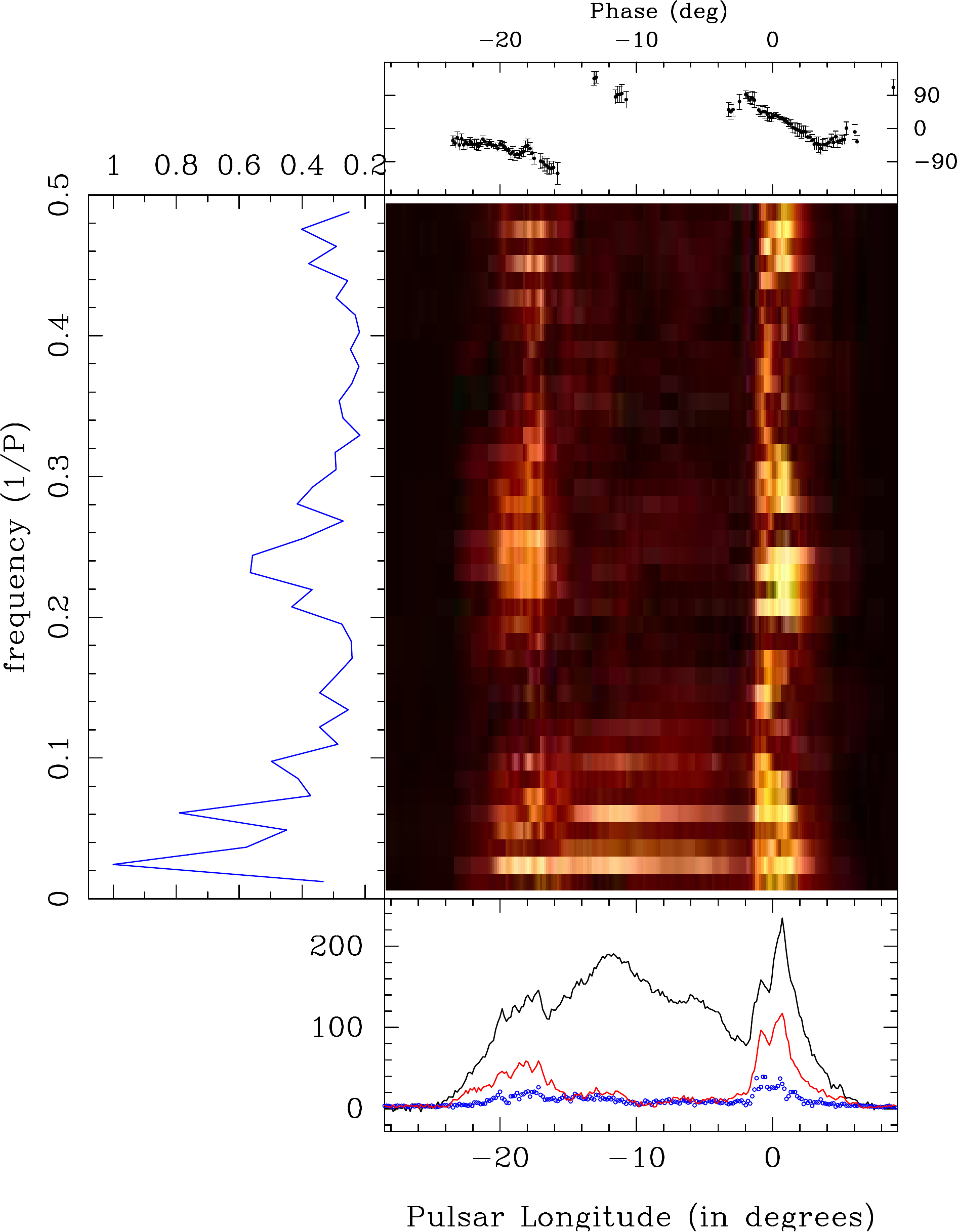}}} \\
{\mbox{\includegraphics[scale=0.4,angle=0.]{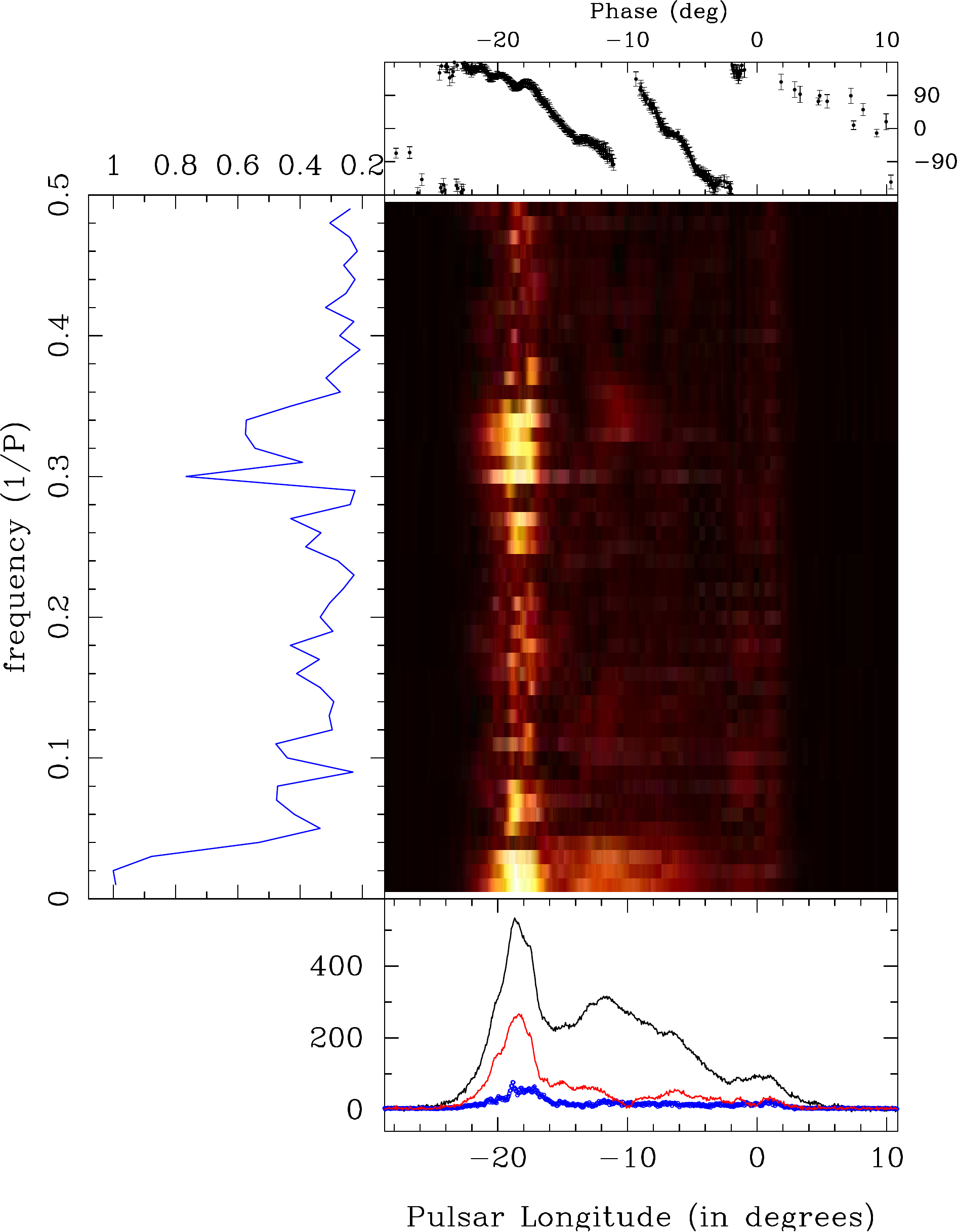}}} &
{\mbox{\includegraphics[scale=0.4,angle=0.]{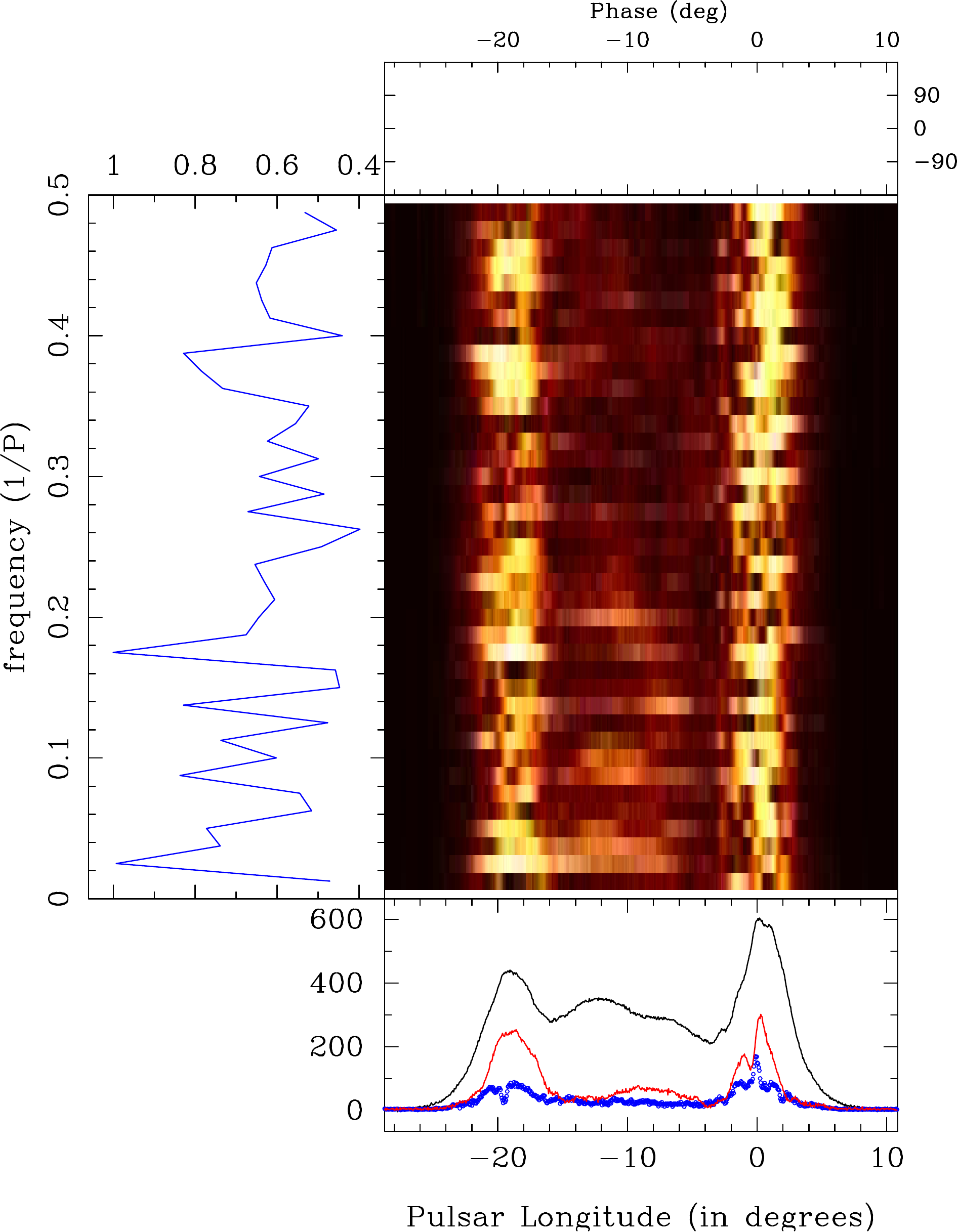}}} \\
\end{tabular}
\caption{The figure shows the longitude resolved fluctuation spectra (LRFS) 
corresponding to four emission modes of PSR J2321+6024. The top left panel 
shows the average LRFS of mode A from four pulse sequences on 4 November 2017. 
The top right panel shows the LRFS between pulse 1722 and 1799 on 16 November 
2018, when the pulsar was primarily in the B mode interspersed will nulls. The 
bottom left panel corresponds to the pulse range between 1 and 99 on 22 July 
2019, where three short duration ABN modes was seen separated by nulls. The 
bottom right panel shows the average LRFS during mode C from two separate 
sequences on 22 July 2019. The central color map shows the variation of the 
amplitude of fluctuation frequency across the pulse window, whose average value
across the window is shown in the left window. The window in the bottom shows 
the profile formed after averaging the pulses used in the LRFS (in black), 
along with the variation of the amplitude of the peak frequency (in red) and 
the baseline rms level of the frequency amplitude (in blue). The top window 
shows the phase variation across the pulsar profile, corresponding to the peak 
drifting frequency in the left window.}
\label{fig:Mode_LRFS}
\end{figure*}

\noindent
 The subpulse drifting behavior in the different emission modes of PSR 
J2321+6024 was investigated using the longitude resolved fluctuation spectra
\citep[LRFS,][]{1973ApJ...182..245B}, where Fourier analysis is carried out across each longitude range within the pulse window, for a suitable number of pulses. The drifting is quantified using the LRFS, where the presence of any feature with peak intensity at a frequency ($f_p$) in the spectrum corresponds to a drifting periodicity ($P_3=1/f_p$) within the pulse sequence. The error in
estimating $f_p$ is obtained from the full width at half maximum (FWHM) of the 
feature such that $\delta f_p$=FWHM/$2\sqrt{2\ln{2}}$ 
\citep{2016ApJ...833...29B}, and $\delta P_3=\delta f_p/f_p^2 $. The Fourier analysis gives a complex spectrum ($I_{\nu}$) where phase behaviour at the peak
frequency illustrates the relative variations of the emission across the pulse window. For example, a 180\degr~phase difference between two longitudes implies that when the emission is seen at maximum intensity in the first place the corresponding intensity in the second is zero and vice versa. The errors in 
the estimating phase are computed as \citep{2019MNRAS.486.5216B}:\\
$\delta \phi= \frac{\displaystyle |x\delta y- y \delta x| }{\displaystyle x^2+y^2}; \quad$ where $\quad x=Re(I_\nu), \quad  y= Im(I_\nu)$, \\
where $\delta x$ and $\delta y$ have been computed from the baseline rms of the
real and imaginary parts of the complex spectra at horizontal bins corresponding to $f_p$.

The presence of multiple emission modes with varying durations during each
observing session makes it challenging to quantify the drifting behaviour. The lack of continuity between different sections of the same mode results in arbitrary phase shifts in the LRFS. Besides, the long observations of 
pulsars with subpulse drifting have shown that in the majority of cases the 
drifting feature is not seen as a narrow peak, but show fluctuations around a 
mean value at timescales of hundreds of periods \citep{2018MNRAS.475.5098B,
2020arXiv200803329B}. However, it is still possible to obtain average spectra from multiple sections of the different emission modes, by aligning the
phase at a specific longitude, usually the profile peak position, before the averaging process. In the case of the longer duration modes A and C, there were multiple sections where the LRFS could be estimated. However, for the short duration modes B and ABN there were a limited number of sections, interspersed with nulls, available for these studies. Fig. \ref{fig:Mode_LRFS} shows 
examples of LRFS for all four modes. The average LRFS from 4 November 2017 is 
shown in the figure (top right panel), where four sequences above 100 
$P$ (see Table \ref{tab:modelist} in Appendix) was used to form the spectra. The LRFS shows a narrow 
drifting peak around $f_p\sim$ 0.1 cy/$P$. In the case of modes B (top right panel)
and ABN (bottom left panel), in addition to the drifting features seen around frequencies of 0.2 cy/$P$ and 0.3 cy/$P$, respectively, low-frequency 
modulations due to periodic nulling are also seen in the fluctuation spectra. 
The bottom right panel shows the average LRFS for mode C where two pulse sequences $>$ 80 $P$ from 22 July 2019 were used. No clear periodic behaviour 
is seen in the LRFS.

The average drifting periodicity during mode A was estimated to be 
7.8$\pm$0.3 $P$, in mode B was 4.3$\pm$0.4 $P$, and in mode ABN was 3.1$\pm$0.2
$P$. Drifting was seen in all four components of the profile during mode A, 
with the outer components showing flatter phase variations compared to the inner components. There were distinct changes in phase behaviour around the component boundaries and nearly 180\degr~phase separation between the first and
fourth components. In mode B, the drifting was seen in the first and the fourth components which were relatively flat with nearly 90\degr~phase 
separation between them. No clear drifting feature was seen in the inner components. Finally, in the case of mode ABN  drifting was negligible towards the trailing edge where the emission vanished. However, unlike mode B, the drifting was prominently seen in the inner components with large phase variations. The phase behaviour was similar to mode A with the first component having relatively flat phase variations and transitions were seen in the phase behaviour around the component boundaries. The third component in mode ABN 
had much steeper phase variations compared to mode A.

\subsection{Emission height and Geometry}\label{Emission height and geometry}

\begin{figure}
\centering
{\mbox{\includegraphics[scale=0.28,angle=0.]{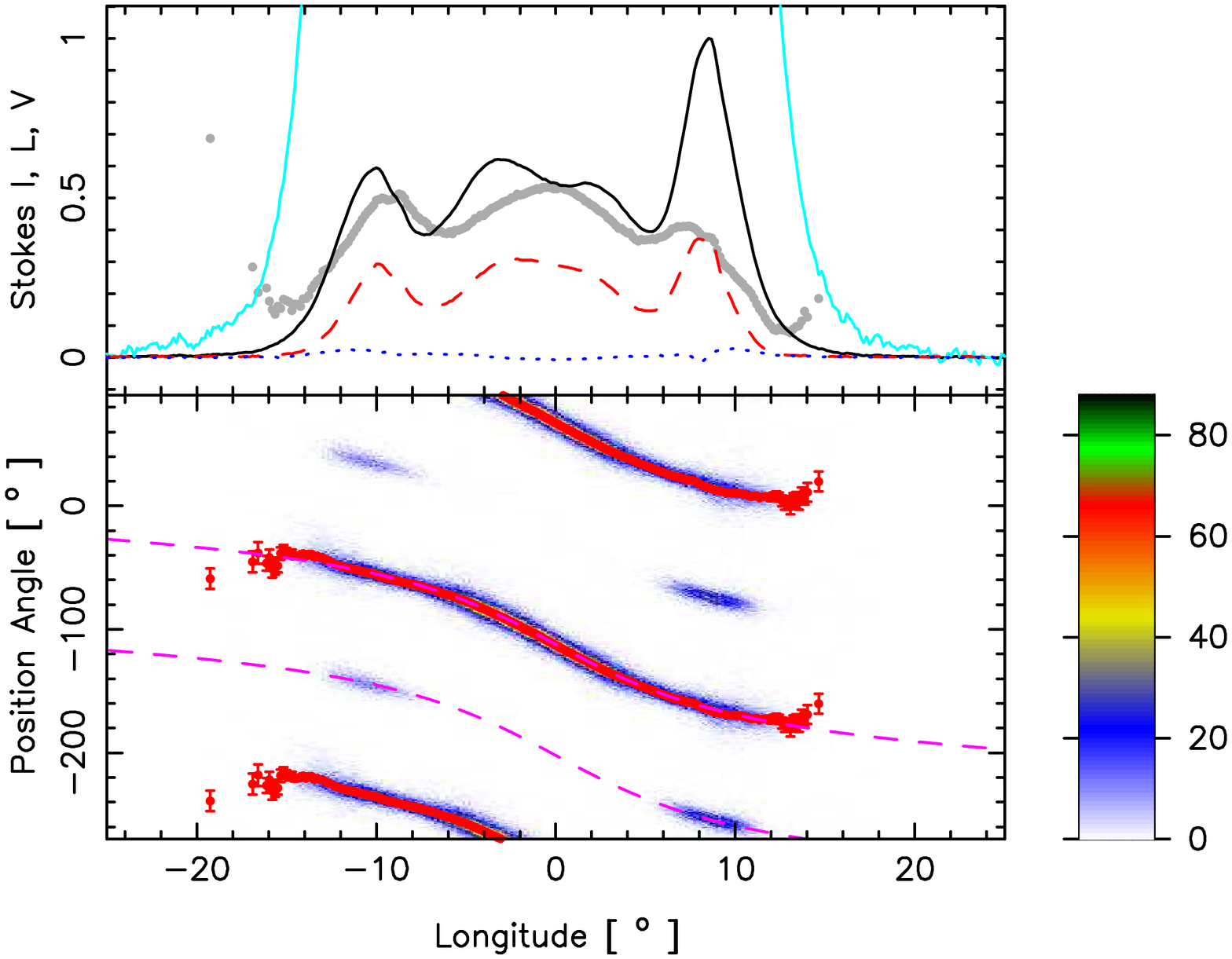}}} \\
\vspace{15px}
{\mbox{\includegraphics[scale=0.28,angle=0.]{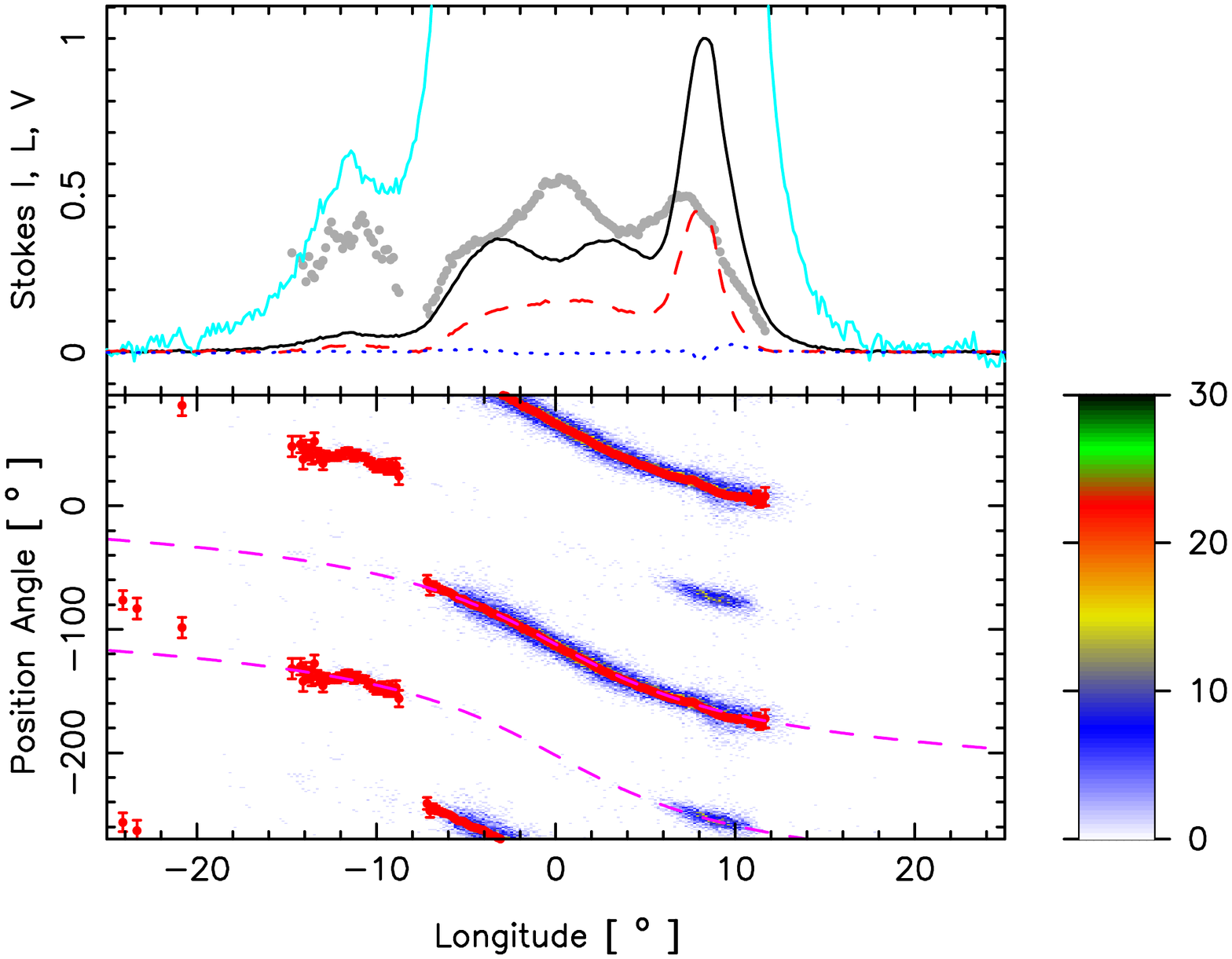}}} \\
\vspace{15px}
{\mbox{\includegraphics[scale=0.28,angle=0.]{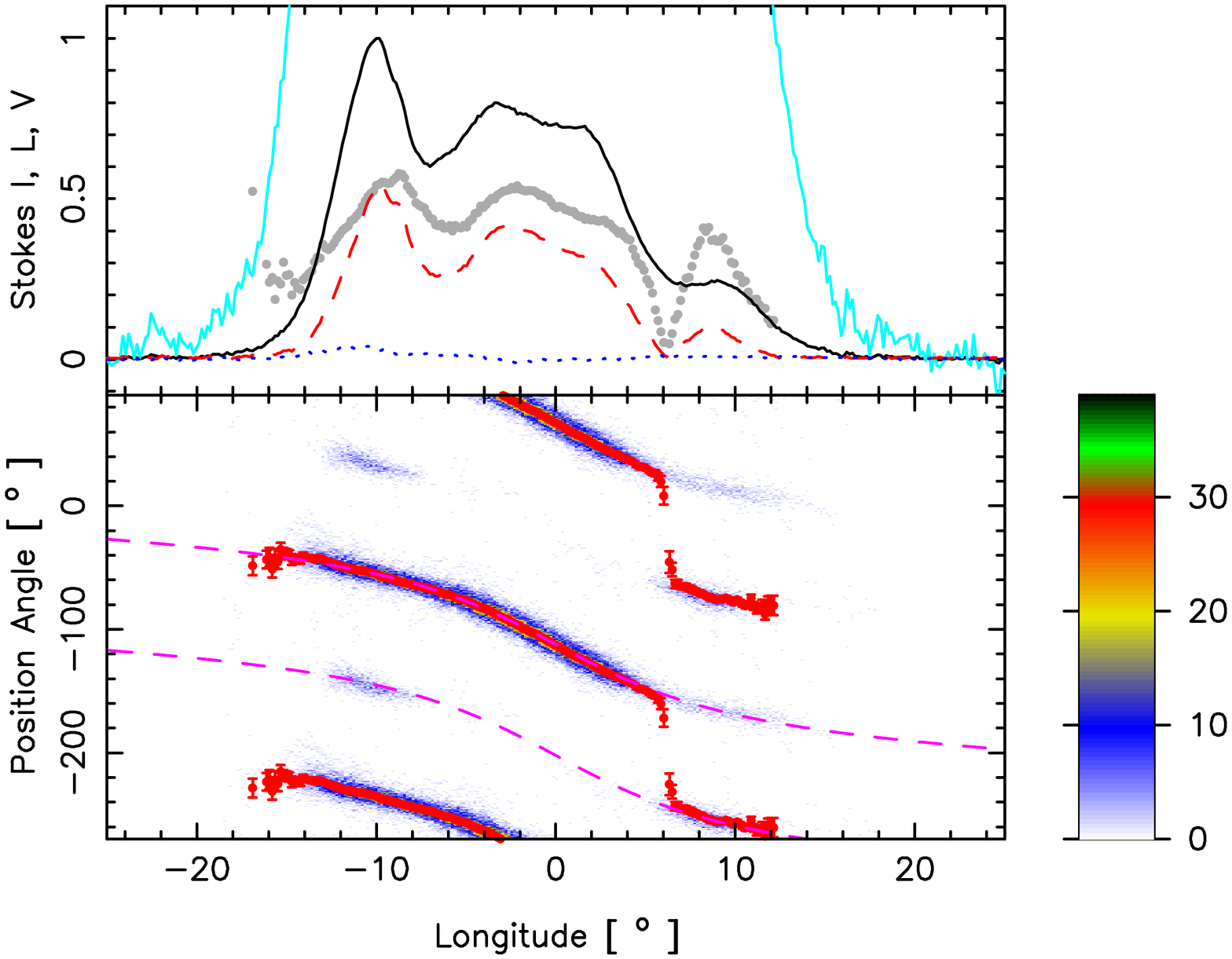}}} \\
\caption{The figure shows the total intensity (Stokes I; solid black), the
average linear (L; dashed red), the circular polarization(Stokes V; dotted 
blue), the fractional linear polarization (L/I; grey points) and a zoomed total
intensity (10$\times$I; cyan curve), for 22 July, 2019 observations full band average from 314.5-474.8 MHz. The PPA histogram (lower window) within each 1\degr$\times$1\degr
sample cell correspond to samples having errors less than 3$\sigma$ in linear 
power are plotted according to the colour bars at the lower right. The average 
PPA traverses (red) are overplotted, and the RVM fit to the PPA traverse is 
plotted twice for the two polarization modes (magenta dashed). The origin is 
taken at the fitted PPA inflection point. The top panel presents the entire 
observing duration, while in the central panel only low signal to noise of 
leading component is chosen and in the bottom plot the low signal to noise 
trailing component is selected.}
\label{fig_polprof}
\end{figure}

\begin{figure}
\centering
{\mbox{\includegraphics[scale=0.34]{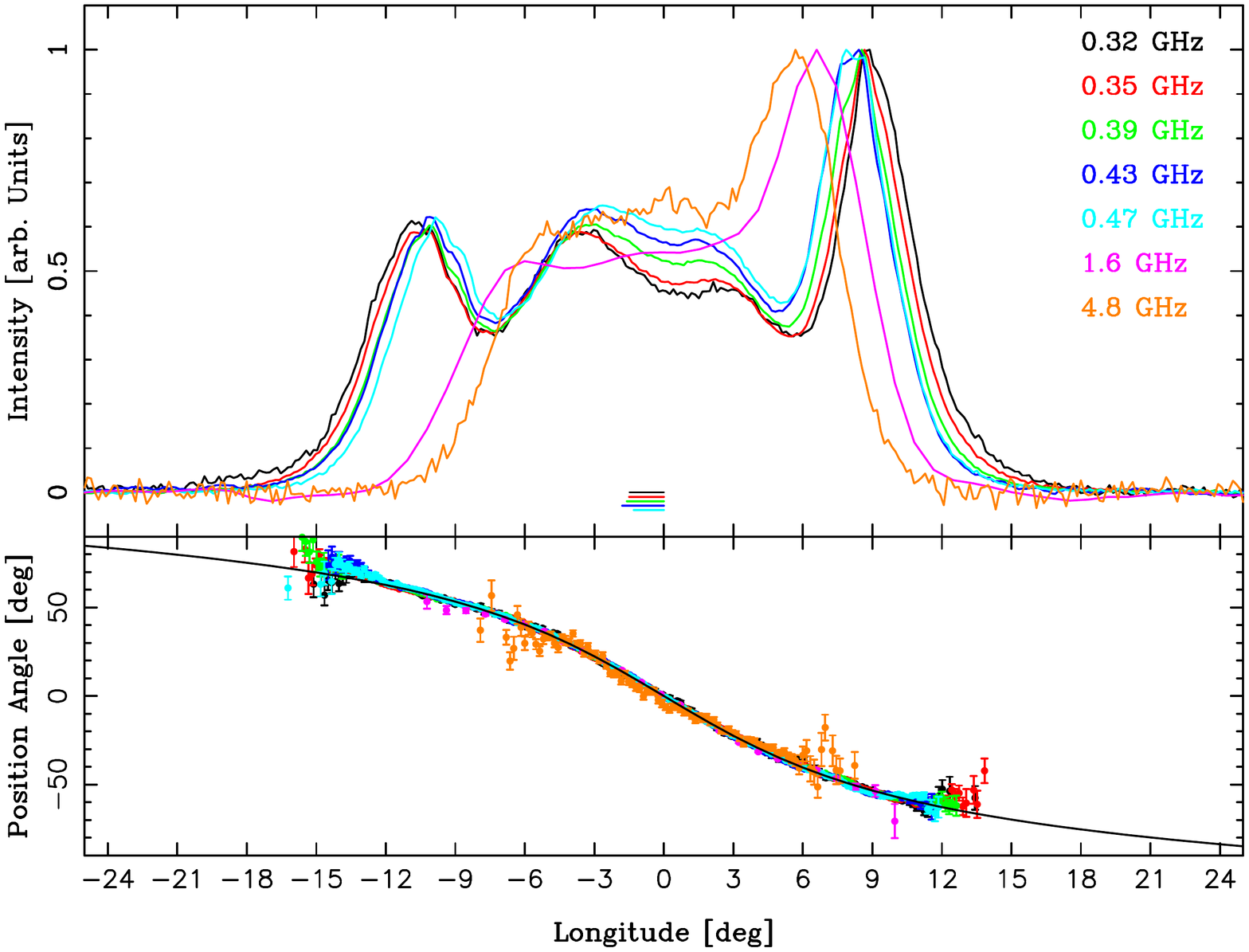}}} \\
\vspace{20px}
{\mbox{\includegraphics[scale=0.34]{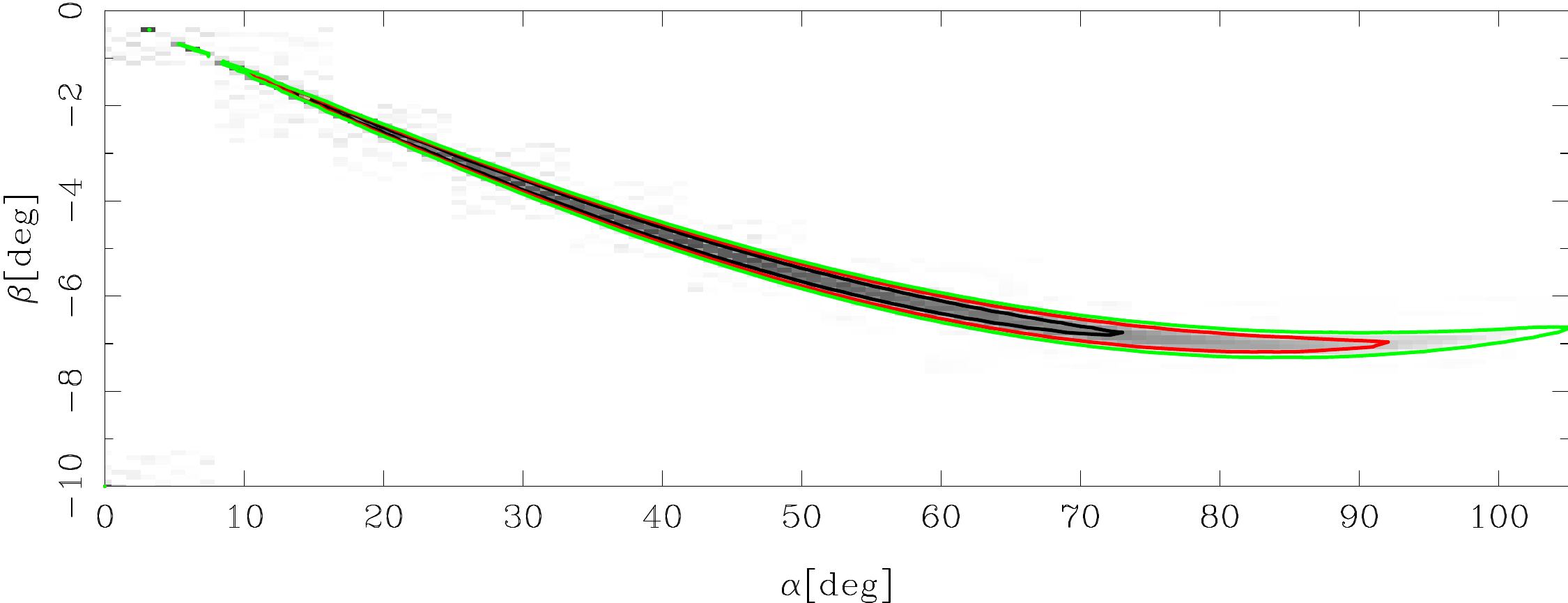}}} \\
\caption{The top window shows the average polarization properties of PSR 
J2321+6024 for five frequency sub-bands between 300 and 500 MHz from the 22 
July, 2019 observations, as well as 1.6 GHz \citep{1998MNRAS.301..235G} and 4.8
GHz \citep{1997A&AS..126..121V}. The top panel shows the total intensity 
profile and the bottom panel the polarization polarization position angle (PPA)
over the wide frequency range. The solid line displays the RVM fit using 
$\alpha =18\degr$ and $\beta=-2.4\degr$. The bottom window shows the chi-squared 
distribution for the fitted parameter $\alpha$ and $\beta$, where the contours 
in black red and green corresponds to 1, 2 and 3 times the minimum chi-squared 
values. The parameters $\alpha$ and $\beta$ are highly correlated in the fit.}
\label{fig_fitPPA}
\end{figure}

PSR J2321+6024 is classified as a conal Quadrupole by 
\citet{1993ApJS...85..145R}. The pulsar profile has four conal components, all of which show subpulse drifting, and is consistent with this classification under the core-cone model of pulsar emission beam. We
have explored the average and single pulse polarization behavior from wideband
observations. The average and single pulse polarization position angle (PPA) is shown in Fig. 
\ref{fig_polprof} (top panel) where two orthogonal tracks, with the primary PPA track existing all along with the profile, and the secondary track mostly existing for the outer components. We used the profile mode separated data for modes A, B, ABN, and C respectively to obtain polarization tracks (not shown) which follow
similar behavior as seen in Fig. \ref{fig_polprof}, i.e. the primary PPA track dominates in each profile mode and orthogonal polarization moding (hereafter OPM) is seen in all the modes. In certain pulsars like PSR B0329+54 and B1237+25 it has been reported that the primary polarization mode is associated with stronger intensity pulses and the secondary mode with weaker pulses (see \citealt{2007MNRAS.379..932M,2013MNRAS.435.1984S}).
To investigate this feature in our data, we used the full pulse sequence and found single pulses 
with low signal to noise or weak emission associated with the leading component within the longitude range -14\degr to -10\degr 
(middle window) and weak emission associated with trailing component lying in the longitude range 6\degr~to 12\degr. The plots in the middle and
the bottom panel of Fig.~\ref{fig_polprof} correspond to the weak leading and trailing emission cases respectively, where 
the polarization flip to the weaker secondary polarization mode is seen. 

We have divided the wide frequency observation into five smaller sub-bands and
estimated the average profile properties for each band. In Table~\ref{tab:Profheight} in Appendix,
the profile widths measured for the various modes A, B, ABN, C, and the grand average profile combining all emission modes referred to as `All' is given. 
The outer profile widths W$_{5\sigma}$ is the profile width measured by identifying points in the leading 
and trailing edge of the profiles that correspond to 5 times the off pulse RMS noise. The W$_{10}$ and W$_{50}$ 
are widths measured by identifying points corresponding to 10\% and 50\% of the peak of the leading and trailing component of the profile. Further, W$^{out}_{sep}$ and W$^{in}_{sep}$
correspond to the peak to peak separation of the outer and inner components respectively. We found that for most cases W$^{out}_{sep}$
could be identified reliably, but W$^{in}_{sep}$ were mostly merged 
hence the peaks could not be identified unambiguously except for mode A. Based on the width
measurements it is clear from Table~\ref{tab:Profheight} that the outer components are consistent 
with the radius to frequency mapping where profile widths are seen to decrease with increasing frequency (see \citealt{2002ApJ...577..322M}). The effect of radius to frequency mapping is clearly 
visible in  Fig.~\ref{fig_fitPPA} (upper panel), where the profiles are seen to 
have progressively lower widths with increasing frequency from 0.3 GHz to 
4.8 GHz which was also noted by \citet{1998MNRAS.299..855K}.

For the average full band collapsed (between  314.5 to 474.8 MHz) data, the percentage of linear polarization fraction is about 37\% and circular polarization fraction is around 1.5\%. We found similar values of percentage of polarization in all the sub-bands. The average
PPA traverse tracks of the primary polarization mode show the characteristic 
S-shaped curve and can be fitted using the rotating vector model (RVM) as 
proposed by \citet{1969ApL.....3..225R}. According to RVM the change in PPA 
($\chi$) traverse reflects the change of the projected magnetic field vector onto the line of sight in the emission region. As the pulsar rotates, the observer samples different 
projection of the magnetic field planes as a function of pulse rotational phase
($\phi$). For a star centered dipolar magnetic field if $\alpha$ is the angle 
between the rotation axis and the magnetic axis, and $\beta$ is the angle 
between the rotation axis and the observer's line of sight on closest approach, then the RVM has a 
characteristic S-shaped traverse given by,

\begin{equation}
\chi = \chi_{\circ} + \tan^{-1} \left( \frac{\sin{\alpha}\sin{(\phi-\phi_{\circ}})} {\sin{(\alpha+\beta)}\cos{\alpha} - \sin{\alpha}\cos{(\alpha+\beta)}\cos{(\phi-\phi_{\circ})}}\right)
\label{eq1}
\end{equation}

Here $\chi_{\circ}$ and $\phi_{\circ}$ are the phase offsets for 
$\chi$ and $\phi$ respectively. At $\chi_{\circ}$ the PPA goes through the steepest gradient (SG) point, which for a static dipole magnetic field is associated with the plane containing the rotation and the magnetic axis. We have used Eq.~\ref{eq1} to reproduce the PPA behavior for the entire frequency
range and the individual sub-bands, and found an excellent fit in all the cases.We also used archival higher frequencies of 1.6 GHz \citep{1998MNRAS.301..235G} and 4.5 GHz
\citep{1997A&AS..126..121V}, and found the RVM fits well to this data.

Further we combine all the PPA traverses following the procedure of 
\citet{2004A&A...421..215M}, where the Eq.~\ref{eq1} was fitted to each PPA 
traverse, and the profiles were aligned after subtracting the off-set 
$\chi_{\circ}$ and $\phi_{\circ}$. The combined PPA traverse was 
fitted to Eq.~\ref{eq1} (shown as solid black line in Fig.~\ref{fig_fitPPA}) 
to obtain $\alpha$ and $\beta$, with offsets set to zero
having errors of $ \delta \phi_{\circ}= \pm 0.1\degr$ and $\delta \chi_{\circ} = \pm 
2\degr$. Although the RVM fit to the combined PPA is excellent, it is difficult to 
determine the geometrical angles because $\alpha$ and $\beta$ are highly 
correlated as seen from the chi-squared contours in Fig.~\ref{fig_fitPPA}, bottom 
window. The difficulty in estimating the pulsar geometry from the RVM fits has 
been highlighted in several earlier studies \citep{2001ApJ...553..341E,
2004A&A...421..215M}. However, the inflexion of the PPA traverse is 
significantly better constrained by observations and depends on the pulsar 
geometry as,
\begin{equation}
\sin(\alpha)/\sin(\beta) 
= \mid d\chi/d\phi \mid_{max} 
\label{eq2}
\end{equation}
Using these estimates we find $\mid d\chi/d\phi \mid_{max} = -7.8\degr \pm0.4 \degr$. 

An alternate route for estimating the geometry has been suggested by using the 
results from the empirical theory (hereafter ET) of pulsar emission as detailed in the 
following studies, ETI \citet{1983ApJ...274..333R}; ETIV 
\citet{1990ApJ...352..247R}; ETVI \citet{1993ApJ...405..285R}; ETVII 
\citet{2002ApJ...577..322M}. According to ET the pulsar emission arises from 
region of dipolar magnetic field at a given height where emission region is 
circular. Thus the conal beam radius $\rho$ is related to the pulsar 
geometrical angles and pulse width $W$ as,
\begin{equation}
\sin^2{\rho/2} = \sin{(\alpha+\beta)}\sin{\alpha}\sin^2{\mathrm{W}/4}+\sin^2{\beta/2}
\label{eq3}
\end{equation}
As argued in ETIV for pulsar having core components, an estimate of $\alpha$ 
could be made since in the ET a relation between the width of the core 
W$_{core}$ at 1 GHz and the angle $\alpha$ could be found as W$_{core} = 2.45\degr \;  
P^{-0.5}/\sin{\alpha}$. Once $\alpha$ is known, Eq.~\ref{eq2} and \ref{eq3} can
be used to find $\beta$ and $\rho$ respectively. Using a large number of 
pulsars with core emission, ETVI established that radius of the inner and outer
conal beams $\rho^{1GHz}_{in/out}$ measured at 1 GHz pulsars follow a simple 
scaling relation with pulsar period, given by $\rho^{1 GHz}_{in/out} = 
4.3^{\circ}/5.7^{\circ} P^{-0.5}$. However, for pulsars like PSR J2321+6024, 
where there is no core emission, we can use $ \beta$ as a function $\alpha$  from Eq.~\ref{eq2}
, the estimate of  $\rho^{1 GHz}_{out}$ and the measured half-power outer width W$_{50}$ at 1 GHz 
and substitute them in Eq. \ref{eq3} to obtain an equation which depends only on $\alpha$.
Further following an iterative procedure and matching the left hand and right hand 
side of Eq. \ref{eq3} we can get estimates of $\alpha$. PSR J2321+6024 has a period of $P$ = 2.256 sec, which 
gives $\rho^{1GHz}_{outer}= 3.8\degr$. Using the 1 GHz half-power width W$_{50}
\sim 10\degr$ the pulsar geometry is estimated to be $\alpha \sim 18\degr$ and 
$\beta \sim -2.4\degr$. We have used this geometry to reproduce the RVM curve 
in Fig.~\ref{fig_fitPPA} (solid black line) which is a good fit to the  combined PPA traverse.

\begin{figure}
\centering
\includegraphics[scale = 0.5]{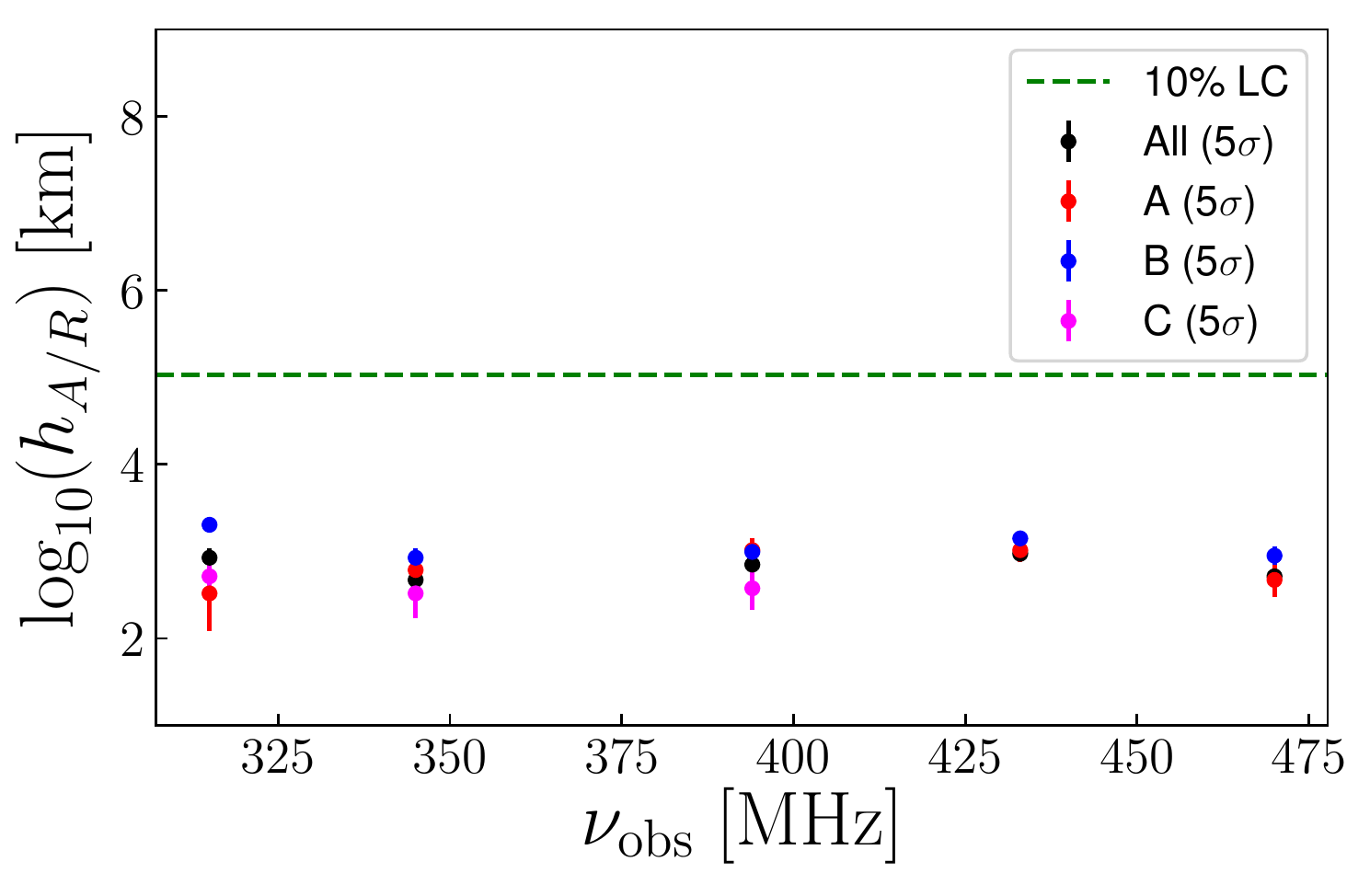}
\caption{The figure shows the estimated emission heights as a function of 
the frequency for the three emission modes A, B, and C as well as the full 
profile for the observations on 22 July 2019. The profile center has been 
obtained from the 10\% level of the peak. The figure also shows the 10\% 
distance to the light cylinder, with all height estimates well below this 
level.}
\label{fig_height}
\end{figure}

The value of $\rho$ also gives the geometrical height of the radio emission,  
$h_{geo} = 10 \; P \left(\rho/1.23\degr\right)^2$, and for PSR J2321+6024 is about 225 
km. This is an approximate estimate which is dependent on several poorly 
constrained parameters as discussed above. RVM is strictly true for a static dipole. A realistic treatment should deal with the contribution of the effects of aberration and retardation as a result of pulsar rotation. In a series of studies by 
\citet{1991ApJ...370..643B,2008MNRAS.391..859D,2020MNRAS.491...80P} it has been
shown that for slow pulsars and radio emission arising below 10\% of the light 
cylinder (hereafter LC), a lag $\Delta \phi$ is observed between the center of the profile and
the steepest gradient or inflection point of the PPA, which is related to the 
radio emission height as, 
\begin{equation}
h_{A/R} = \frac{c P \Delta \phi}{1440}~~ \mathrm{km}
\label{eq3_1}
\end{equation}
where $c$ is the velocity of light. Finding $\Delta \phi$ in Eq.
\ref{eq3_1} requires measurement of the steepest gradient point, which was 
obtained from fitting the RVM, as well as a center of the profile, which 
requires measuring two representative locations $\phi_l$ and $\phi_t$, near the
leading and trailing edges that are equidistant from the center. 
\citet{2004A&A...421..215M} highlighted that the accuracy of identifying 
$\phi_l$ and $\phi_t$ significantly affects the emission height estimates. We 
have used the five frequency sub-bands of 22 July 2019 observations to find 
$h_{A/R}$ in the different emission modes and the average profile, shown in 
table \ref{tab:Profheight}.

The reference locations $\phi_l^{5\sigma}$ and 
$\phi_t^{5\sigma}$ were obtained from the 5$\sigma$ intensity points of the outer conal components 
(see Table~\ref{tab:Profheight} in the Appendix)
of the integrated profile of individual modes A, B, ABN, C, and All combined respectively. 
These were then used to estimate $\Delta \phi$ and emission height (h$_{A/R}$). 
For mode A, B, and All there were sufficient pulses to obtain a reasonably high
signal to noise for the five sub-band data, and thus robust
estimates of emission heights were possible. For modes ABN and C, the signal to noise 
of the profile were not always high to get reliable estimates of  $\phi_{\circ}$, $\phi_l^{5\sigma}$ and 
$\phi_t^{5\sigma}$, and hence the A/R estimated height are prone to larger uncertainties
as seen in Table~\ref{tab:Profheight} in the Appendix. The Table shows that for mode ABN none of the estimated heights are reliable while for mode C only the first three sub-bands have reliable estimates of emission height. 
The emission height for All modes combined and the individual modes across the five sub-bands are very similar and are plotted in Fig.~\ref{fig_height}. We can conclude that emission modes A, B, and C at any given frequency arises from similar heights. In all cases, the emission heights are well below 10\% of the light cylinder
radius.

\section{Discussion} \label{sec:disc}
Using highly sensitive wideband GMRT observations we performed detailed single pulse analysis for PSR J2321+6024 and found four emission modes. We confirm three emission modes associated with subpulse drifting viz., modes A, B, and ABN  as identified by WF81. We also report the presence of a possible new non-drifting mode C. We have characterized the nulling and drifting and polarization properties for each of these emission modes. We find statistical evidence of the evolution of the modal abundance and the average length of the emission modes between the first and the last two runs of our observations.

PSR J2321+6024 falls in the category of pulsars showing the
phenomenon of mode changing associated with a change in drift
periodicity along with nulling. Only a handful of pulsars
(about 9) are known to show
this phenomena, and the physical origin of these phenomena is
poorly understood. However, a comprehensive observational study, as
presented in this work, allow us to draw certain inferences which we discuss below.

Several lines of studies suggest that coherent radio emission from normal pulsars arises due to the growth of 
plasma instabilities in relativistically flowing pair plasma along the open dipolar magnetic field lines. 
Recent observational evidence (e.g.\citealt{2009ApJ...696L.141M}) have shown that pulsar radio emission is 
excited due to coherent curvature radiation (hereafter CCR) and the emission detaches from the pulsar
magnetosphere in regions below 10\% of the LC (see \citealt{2017JApA...38...52M}).
A possible mechanism by which CCR can be excited in pulsar plasma is due to the formation of Langmuir charge solitons, or charge bunches
\citep{2000ApJ...544.1081M,2004ApJ...600..872G,2018MNRAS.480.4526L}. Pulsar plasma is formed due to sparking discharges in the IAR at the polar cap. 
Two models of sparking discharges are available in the literature viz, the purely vacuum RS75 model and a variant of the RS75 model called the Partially Screened Gap (hereafter PSG) model by  \citet{2003A&A...407..315G} (hereafter GMG03). The PSG model allows a thermionic emission of ions from the polar cap. The flow of ions gives rise to an ion charge density $\rho_\mathrm{ion}$ in the polar gap such that the purely vacuum gap potential of RS75 is replaced by the screened potential $\Delta V_\mathrm{PSG} = \eta_\mathrm{PSG} \; \Delta V_\mathrm{vac} $ where the screening factor $\eta_\mathrm{PSG} = 1 - \rho_\mathrm{ions} / \rho_\mathrm{GJ}$ such that $\rho_\mathrm{ions} / \rho_\mathrm{GJ} \approx \exp 30(1 - T_\mathrm{ion} / T_\mathrm{s}) $ where $\rho_\mathrm{GJ}$ is the co-rotational Goldreich-Julian charge density (\citealt{1969ApJ...157..869G}), $T_\mathrm{ion}$ is the critical temperature for which the  screening is complete and $T_\mathrm{s}$ corresponds to the thermal temperature of the polar cap. $T_\mathrm{s} = T_\mathrm{ion} (1 - \delta)$ such that $\delta << 1$ . The critical temperature T$_\mathrm{ion}$ is determined completely by the strength of surface magnetic field. The surface field strength is determined by  the parameter $b$ (see Eq.~3 of GMG03). It is the ratio of the local surface magnetic field strength to the global dipole field strength (see e.g. \citealt{2007MNRAS.382.1833M} ;  \citealt{2013arXiv1304.4203S} ; \citealt{2015MNRAS.447.2295S}). For $b=1$, the surface magnetic geometry is purely dipolar  while $b>1$ corresponds to a non-dipolar surface magnetic field configuration. The surface temperature $T_\mathrm{s}$ of the polar cap is maintained at a quasi-equilibrium value by thermostatic regulation. The thermostatic regulation is such that $T_\mathrm{s}$ can differ from $T_\mathrm{ion}$ by at most $\delta \sim 0.01$ and any greater offset is corrected on timescales of around 100 nanoseconds. Due to this finely tuned regulation of $T_\mathrm{s}$ , the value of $\eta_\mathrm{PSG}$ is maintained at a near steady value. Many recent studies point to the presence of non-dipolar surface magnetic field configuration while the radio emission region has dipolar magnetic field configuration(see e.g. \citealt{2019MNRAS.489.4589A} , \citealt{2020MNRAS.491...80P}). The PSG model combined with non-dipolar surface field configuration can successfully explain a plethora of observations viz. , sub-pulse drifting and thermal X-ray emission  from polar caps (see e.g. \citealt{2003A&A...407..315G, 2015MNRAS.447.2295S, 2016ApJ...833...29B, 2020MNRAS.492.2468M, 2013arXiv1304.4203S} ). For this pulsar PSR J2321+60, GMG03 estimated $b= 13.6$,   $\eta_\mathrm{PSG} =  0.071$, $T_\mathrm{ion} = 1.239 \times 10^6$ K and $T_\mathrm{s} = 1.236 \times 10^6$ K.

We consider the model of the surface non-dipolar magnetic field as
suggested by \citealt{2002A&A...388..235G} (hereafter GMM02). In this model magnetic
field configuration is a superposition of a large
scale neutron star centered global dipolar magnetic
field and a small scale crust anchored dipolar magnetic field. As seen from fig. 2 of GMM02, at the radio emission zone, the magnetic field
is dominated by the global magnetic field parameters, while
the neutron star surface magnetic field is dominated by the crust anchored dipole. In this model $\alpha_\mathrm{l}$ is the angle the local surface magnetic field vector at the polar cap ($\Vec{B}_\mathrm{l}$) makes with the rotation axis ($\vec{\Omega}_\mathrm{Rot}$) and $\alpha$ which is the angle between the global dipole moment $\vec{\mu}_\mathrm{d}$ and $\vec{\Omega}_\mathrm{Rot}$ (inferred from RVM) at the radio emission zone. In GMM02 the radius of curvature at the surface ($\rho_\mathrm{c,l}$) is dominated by the crust anchored dipole while the radius of curvature at the radio emission region ($\rho_\mathrm{c}$) has a dipolar character. Thus in this model by GMM02 the surface magnetic field parameters $\alpha_\mathrm{l}$ and $\rho_\mathrm{c,l}$ can change without changing the magnetic field parameters at the radio emission region.

In the subsequent subsections, we attempt to understand the connections between the different phenomenon of mode changing, nulling, and subpulse drifting based on the PSG model. 

\subsection{Mode emission regions and Plasma conditions}

The excellent fit of the PPA to the RVM
for a wide frequency range is evident from Fig.~\ref{fig_fitPPA}.
This suggests that the radio emission originate from regions of open dipolar
magnetic field lines.
As shown in section~\ref{Emission height and geometry}, the application of the A/R method
to the polarization data across a wide
frequency range gives estimates of emission heights to be below
10\% of the light cylinder radius for the modes A, B, and C. As mentioned earlier the estimates of height for mode ABN were unreliable. From
Fig \ref{fig_height} and table~\ref{tab:Profheight} it is clear
that the emission from the modes A, B, and C arises
from very similar heights. Such detailed A/R analysis has been
applied to only other mode changing pulsar PSR B0329+59 (see 
\citealt{2019MNRAS.484.2725B}), where
a similar conclusion was reached, viz. various emission modes arise from similar emission height.

We follow arguments similar to \citealt{2019MNRAS.484.2725B} to understand the implication of similar emission heights for various emission modes. We conclude that similar emission regions for modes A, B, and C implies a similar radius of curvature $\rho_\mathrm{c}$ for these modes. The coherence in CCR is achieved at the critical frequency $\nu_\mathrm{c}$ given by

\begin{equation}
    \nu_\mathrm{c} = \frac{3}{4 \pi} \frac{\gamma^3 c}{\rho_\mathrm{c}}
    \label{critical frequency}
\end{equation}
where $\gamma$ is the Lorentz factor of charge solitons. We assume the observing frequency ($\nu_\mathrm{obs}$) at which coherent radio emission is received as being equal to the critical frequency ($\nu_\mathrm{c}$). Since $\nu_\mathrm{obs}$  and $\rho_\mathrm{c}$  are similar, from Eq. \ref{critical frequency} we infer $\gamma$ for all these emission modes to be similar as well. The power due to CCR at $\nu_\mathrm{c}$ is
 \begin{equation}
    P_\mathrm{CCR} = \frac{f(Q^2) \gamma^4 c}{\rho^2_\mathrm{c}}
    \label{power due to CCR}
\end{equation}
where $f(Q^2)$  is a quantity with dimensions of charge squared.  
Since $\gamma$ and $\rho_\mathrm{c}$ are similar, the difference in observed power in these emission modes is due to different $f(Q^2)$. As can be seen from  Fig. \ref{fig_modeprof} for the trailing conal components the ratio of $f(Q^2)$ in modes A, B, and C is different. $f(Q^2)$ in turn, depends on the very complex underlying plasma processes. One way to change to f($Q^2$) is to change the separation of the electron positron distribution function in the pair plasma. Recently, \citealt{2020MNRAS.497.3953R} used the model by GMM02 and showed that this separation at the emission region can be changed by changing $\cos \alpha_\mathrm{l}$ at the surface while keeping $b$ constant. \citealt{2015MNRAS.447.2295S} alluded to two channels by which pair creation can proceed in the PSG model and postulated that different modes are associated with these different channels of pair creation.

In our observations, we also find the presence of OPMs under the leading and the trailing components of the profile. In general, the OPM correspond to the extraordinary (X) and ordinary (O) eigenmodes of the plasma, however in the case of PSR 
J2319+6024 we do not have any means to associate the PPA tracks with X or O mode. It must be noted that the secondary PPA track, which is present primarily below the trailing and leading outer component, dominates when the emission is weaker. Such
an effect has also been reported for PSR B0329+54 wherein the weaker mode could be identified as the O mode (see \citealt{2007MNRAS.379..932M}). Incidentally, modes A and B are associated with weaker emission states associated with the leading component while mode ABN is associated with a weaker emission state with the trailing component.  It must be
noted that the PPA behavior for the modes A, B, 
ABN, and C are very similar and no mode dependent changes are evident.

\subsection{Drift periodicity during Mode changing}

The observed drift periodicity for modes A, B, and ABN is given by $\approx$ 7.8 $P$, 4.3 $P$, and 3.1 $P$ respectively. We find no evidence of drifting in mode C. 
Generally, the observed drift rate (P$_\mathrm{3, obs}$) can either be the true drift rate P$_\mathrm{3, True}$ (for complete Nyquist sampling P$_\mathrm{3, True} > 2 P$  ) or an aliased drift rate P$_\mathrm{3, Aliased}$ (for incomplete Nyquist sampling P$_\mathrm{3, True}< 2 P$). Thus,  in the absence of aliasing we have P$_\mathrm{3,obs}$ = P$_\mathrm{3, True}$  while in the presence of aliasing, the observed drift rate is given by  P$_\mathrm{3, obs}$ = P$_\mathrm{3, Aliased} = $P$_\mathrm{3, True}$/(P$_\mathrm{3, True}$ - 1) (in units of $P$). Assuming aliasing, P$_\mathrm{3, True}$  for modes A, B, ABN and C are given by 1.15 $P$, 1.30 $P$, 1.48 $P$ and $\sim$ 1 $P$ respectively. In the PSG model,the drift periodicity is given by  P$_\mathrm{3, PSG} =  1 / 2 \pi  \eta_\mathrm{PSG} \cos \alpha_\mathrm{l}$ (in units of $P$). As already mentioned before $\eta_\mathrm{PSG}$ can be assumed to remain constant during mode changing. If all the emission modes seen are aliased then for modes A, B and ABN we have  $ \cos {\alpha}_\mathrm{A,l} :  \cos {\alpha}_\mathrm{B,l} : \cos {\alpha}_\mathrm{ABN,l} \approx 87: 77: 68$. In the aliased scenario the drift periodicity for mode C has to be close to 1 $P$ which can be achieved for $\alpha_\mathrm{l} \rightarrow 0$. Next we consider the scenario where the observed drift periodicity is non-aliased. GMG03 estimated $\eta_\mathrm{PSG} = 0.071$ for PSR J2321+60. This gives minimum drift periodicity P$_\mathrm{3,min} = 2.2 > 2 P $. For this value of $\eta_\mathrm{PSG}$, aliasing effects can be discarded and we have P$_\mathrm{3, obs}$ = P$_\mathrm{3, True}$. Under this assumption we have for modes A, B, and ABN we have $ \cos {\alpha}_\mathrm{A,l} :  \cos {\alpha}_\mathrm{B,l} : \cos {\alpha}_\mathrm{ABN,l} = 13: 23: 32$. In the non-aliased scenario, P$_\mathrm{3, PSG}$ for mode C needs to be significantly large. This can be achieved for $\alpha_\mathrm{l} $ approaching $ \pi/ 2 $. Assuming P$_\mathrm{3, obs}$ = P$_\mathrm{3, True}$,
\cite{2016ApJ...833...29B}, \cite{2017ApJ...846..109B} , \cite{2020ApJ...889..133B} found an anti-correlation between P$_{3}$ and $\dot{E}$ , found sub-pulse drifting to be associated with $\eta_\mathrm{PSG} = 0.1$ and $\dot{E} < 10^{33}$ ergs/s. For our pulsar PSR J2321+6024 we have $\dot{E} \sim 10^{31}$ ergs/s and is consistent with this correlation. It must be noted that for the aliased scenario the variation in $\cos{\alpha}_{l}$ from one mode to another is smaller as compared to the non-aliased scenario.

Recently, \cite{2020MNRAS.496..465B} showed that the large phase variation within the profile from one emission mode to the other (as shown in Fig.~\ref{fig:Mode_LRFS}) can be understood due to a changing surface non-dipolar magnetic field structure. Thus, the PSG model can explain the change of drift periodicity and phase variation in the emission profile during mode changing using a variation of surface non-dipolar magnetic field structure.

Finally as seen in the previous subsection, the changes in $\eta_\mathrm{PSG} \cos \alpha_\mathrm{l}$ also affect $f(Q^2)$ (and hence the variation of power in the modes). Thus, a connection between mode change and drifting can be inferred.

\subsection{Nulling and mode changing}

 Table \ref{tab:nullstat} shows that the null state corresponds to a  cessation of any detectable radio emission. The suppression of emission is broad-band.  Two broad suggestions are available in the literature for the null state viz., the change in viewing geometry such that the beam does not cross LOS ( see e.g. \citealt{2010MNRAS.408L..41T})  or the loss of coherence of radio emission (see e.g. \citealt{1982ApJ...263..828F}). The first suggestion requires a pan magnetosphere reconfiguration as the reorientation of the emission beam requires rearrangement of the open field lines. The other alternative consists of plasma generation due to sparking in the polar gap and the onset of two-stream instability in the plasma clouds a few hundreds of km from the surface. The plasma generation process cannot be halted as there are enough hard-gamma ray photons in the diffuse gamma-ray background to restart sparks on timescales of a few 10-100 $\mu$s (see \citealt{1982ApJ...258..121S}). We suggest nulling to be due to broad-band suppression of two-stream instabilities at the radio emission heights. Recently, \citealt{2020MNRAS.497.3953R} used the model by GMM02 and showed that broad-band suppression of plasma instabilities (and hence radio emission) can be achieved by changing non-dipolar surface magnetic field configurations.

Using the PSG model, we put forward the following conjecture that the nulls correspond to values of $\eta_\mathrm{PSG} \cos \alpha_\mathrm{l}$ for which two-stream instabilities are suppressed and the three emission modes correspond to unique $\eta_\mathrm{PSG} \cos \alpha_\mathrm{l}$  for which solitons can form with high but different values of $f(Q^2)$. This can explain the abrupt nature of emission modes as well as a continuous subpulse drifting state. In the PSG model, the nulls can be interpreted as broad-band non-emission mode(s) which are encountered while changes occur in surface ion flow and the radius of curvature of the magnetic field lines at the surface. Additionally, the quasi-periodicity of the nulls points to a quasi-periodicity of this product as well.

\section{Summary \& Conclusions} 
\label{sec:sum}

We have characterized moding, nulling, and drifting features for PSR J2321+6024. The data for three observing runs have been analyzed. This pulsar shows 
four-component cQ morphology. 
Four emission mode exists identified as modes A, B, ABN (as was found by WF81) and a possible new mode C. Modes A , B and ABN are characterized by subpulse drift rates of 7.8$\pm$0.3 $P$ , 4.3$\pm$0.4 $P$ and 3.1$\pm$0.2 $P$ respectively. We find no evidence of subpulse drifting in mode C. The nulls have a quasi-periodic behavior with periodicity 100-150 $P$. The polarimetric analysis shows that the different modes arise from similar emission heights. The mode changes were also accompanied by OPMs. The modal abundance and the average length of the different modes show evolution across three observing runs. We show that the PSG model is a very promising model for interpreting the connection between mode changing, nulling, and sub-pulse drifting.

However, it is difficult to understand from the PSG model alone as to how these distinct drift rates are selected. Surface physics plays a crucial role in the PSG model connecting mode changing to sub pulse drifting and nulling. The broad-band nature of these three phenomena can be understood using the PSG model only if surface physics ( i.e., ion flow and surface magnetic field configuration) can be connected to soliton formation across a range of emission heights.

\section*{Acknowledgements}
We thank the anonymous referee for  critical comments and suggestions that greatly improved the quality of the manuscript. We thank the staff of the GMRT who have made these observations
possible.  The GMRT is run by the National Centre for Radio
Astrophysics of the Tata Institute of Fundamental Research.  Sk. MR
and DM acknowledge the support of the Department of Atomic Energy,
Government of India, under project no. 12-R\&D-TFR-5.02-0700. DM
acknowledges support and funding from the ‘Indo-French Centre for the
Promotion of Advanced Research - CEFIPRA’ grant IFC/F5904-B/2018.

\section*{Data availability}

The data underlying this article will be shared on reasonable request to the corresponding author, Sk. Minhajur Rahaman. 




\bibliographystyle{mnras}
\bibliography{References}



\appendix

\begin{table*}
\centering
\caption{List of the pulse range and duration (Len) of the four emission modes, A, B, ABN and C during each observing session.}
\label{tab:modelist}
\begin{tabular}{c@{\hskip5pt}c@{\hskip5pt}c@{\hskip18pt}c@{\hskip5pt}c@{\hskip5pt}c@{\hskip5pt}c@{\hskip5pt}c@{\hskip5pt}c@{\hskip18pt}c@{\hskip5pt}c@{\hskip5pt}c@{\hskip5pt}c@{\hskip5pt}c@{\hskip5pt}c}
\hline
\multicolumn{3}{c}{4 Nov, 2017} & \multicolumn{6}{c}{16 Nov, 2018} & \multicolumn{6}{c}{22 July, 2019} \\
\hline
Pulse & Mode & Len & Pulse & Mode & Len & Pulse & Mode & Len & Pulse & Mode & Len & Pulse & Mode & Len \\
(P) &  & (P) & (P) &  & (P) & (P) &  & (P) & (P) &  & (P) & (P) &  & (P) \\
\hline
   14 - 116  &  A  & 103 &    1 - 10   &  B  & 10 & 1418 - 1425 &  B  & ~8 &    1 - 35   & ABN & 35 & 1477 - 1481 & ABN & ~5 \\
  120 - 127  &  B  & ~~8 &   17 - 26   &  B  & 10 & 1426 - 1438 & ABN & 13 &   51 - 68   & ABN & 18 & 1496 - 1508 &  B  & 13 \\
  129 - 134  &  C  & ~~6 &   27 - 34   &  C  & ~8 & 1443 - 1530 &  A  & 88 &   84 - 99   & ABN & 16 & 1519 - 1530 &  B  & 12 \\
  155 - 184  &  C  & ~30 &   65 - 72   &  C  & ~8 & 1538 - 1552 &  B  & 15 &  111 - 162  &  A  & 52 & 1537 - 1543 &  B  & ~7 \\
  187 - 207  & ABN & ~21 &   77 - 103  &  A  & 27 & 1555 - 1560 &  C  & ~6 &  177 - 195  &  A  & 19 & 1544 - 1548 & ABN & ~5 \\
  231 - 259  &  A  & ~29 &  152 - 169  &  C  & 18 & 1561 - 1574 & ABN & 14 &  197 - 208  &  B  & 12 & 1558 - 1586 &  A  & 29 \\
  264 - 315  &  C  & ~52 &  181 - 269  &  A  & 89 & 1589 - 1634 &  C  & 46 &  221 - 230  & ABN & 10 & 1596 - 1606 &  A  & 11 \\
  332 - 344  &  A  & ~13 &  272 - 303  &  B  & 32 & 1656 - 1666 &  C  & 11 &  240 - 250  &  A  & 11 & 1708 - 1721 &  B  & 14 \\
  370 - 385  & ABN & ~16 &  309 - 317  &  B  & ~9 & 1674 - 1683 &  B  & 10 &  251 - 257  & ABN & ~7 & 1731 - 1754 &  A  & 24 \\
  388 - 398  &  C  & ~11 &  318 - 328  & ABN & 11 & 1692 - 1720 &  A  & 29 &  265 - 273  &  A  & ~9 & 1757 - 1787 &  B  & 31 \\
  403 - 463  &  A  & ~61 &  349 - 389  &  A  & 41 & 1722 - 1734 &  B  & 13 &  284 - 303  &  A  & 20 & 1790 - 1794 & ABN & ~5 \\
  469 - 477  &  B  & ~~9 &  401 - 429  &  C  & 29 & 1758 - 1776 &  B  & 19 &  327 - 363  &  A  & 37 & 1811 - 1815 & ABN & ~5 \\
  496 - 577  &  A  & ~83 &  447 - 478  &  A  & 32 & 1790 - 1799 &  B  & 10 &  374 - 383  &  A  & 10 & 1838 - 1848 &  B  & 11 \\
  585 - 601  &  C  & ~17 &  487 - 497  &  A  & 11 & 1803 - 1808 & ABN & ~6 &  384 - 391  & ABN & ~8 & 1849 - 1860 &  A  & 12 \\
  604 - 624  & ABN & ~21 &  503 - 511  & ABN & ~9 & 1837 - 1849 &  A  & 13 &  439 - 447  &  A  & ~9 & 1902 - 1981 &  A  & 80 \\
  658 - 804  &  A  & 146 &  546 - 559  &  B  & 14 & 1856 - 1865 &  C  & 10 &  455 - 466  &  A  & 12 & 1989 - 2018 &  B  & 30 \\
  810 - 820  &  B  & ~11 &  562 - 573  & ABN & 12 & 1880 - 1890 &  B  & 11 &  477 - 500  &  B  & 24 & 2026 - 2034 &  B  & ~9 \\
  839 - 850  &  B  & ~12 &  631 - 644  &  B  & 14 & 2013 - 2095 &  C  & 83 &  501 - 517  &  C  & 17 & 2062 - 2085 &  B  & 24 \\
  854 - 862  & ABN & ~~9 &  645 - 657  & ABN & 13 & 2096 - 2110 &  B  & 15 &  528 - 537  &  C  & 10 & 2088 - 2127 &  A  & 40 \\
  896 - 979  &  A  & ~84 &  674 - 740  &  A  & 67 & 2115 - 2129 & ABN & 15 &  566 - 649  &  A  & 84 & 2134 - 2140 &  B  & ~7 \\
  980 - 1003 &  C  & ~24 &  747 - 757  &  B  & 11 & 2150 - 2180 &  C  & 31 &  662 - 674  &  B  & 13 & 2162 - 2191 &  A  & 30 \\
 1042 - 1081 &  C  & ~40 &  761 - 768  & ABN & ~8 & 2201 - 2221 &  A  & 21 &  678 - 686  & ABN & ~9 & 2192 - 2208 &  B  & 17 \\
 1117 - 1138 &  C  & ~22 &  776 - 788  &  A  & 13 & 2229 - 2241 &  B  & 13 &  704 - 714  &  C  & 11 & 2231 - 2267 &  A  & 37 \\
 1175 - 1241 &  A  & ~67 &  791 - 809  &  B  & 19 & 2242 - 2250 & ABN & ~9 &  732 - 777  &  A  & 46 & 2284 - 2299 & ABN & 16 \\
 1248 - 1264 &  B  & ~17 &  813 - 820  & ABN & ~8 & 2257 - 2265 &  C  & ~9 &  778 - 791  &  B  & 14 & 2318 - 2327 &  B  & 10 \\
 1267 - 1337 &  C  & ~71 &  860 - 883  &  A  & 24 & 2286 - 2372 &  A  & 87 &  800 - 817  &  C  & 18 & 2330 - 2339 & ABN & 10 \\
 1358 - 1389 &  C  & ~32 &  892 - 904  &  A  & 13 & 2375 - 2390 &  B  & 16 &  828 - 845  &  A  & 18 & 2348 - 2357 & ABN & 10 \\
 1431 - 1443 &  A  & ~13 &  912 - 926  &  B  & 15 & 2406 - 2412 &  C  & ~7 &  852 - 869  &  B  & 18 &             &     &    \\
 1468 - 1475 &  C  & ~~8 &  927 - 938  & ABN & 12 & 2413 - 2425 & ABN & 13 &  877 - 962  &  A  & 86 &             &     &    \\
 1481 - 1499 &  B  & ~19 &  945 - 1032 &  A  & 88 & 2430 - 2440 &  A  & 11 &  973 - 1053 &  C  & 81 &             &     &    \\
 1521 - 1646 &  A  & 126 & 1033 - 1044 &  B  & 12 & 2448 - 2454 &  C  & ~7 & 1062 - 1070 & ABN & ~9 &             &     &    \\
 1647 - 1660 &  B  & ~14 & 1073 - 1086 &  B  & 14 & 2474 - 2543 &  A  & 70 & 1086 - 1102 &  A  & 17 &             &     &    \\
 1662 - 1682 &  C  & ~21 & 1087 - 1099 & ABN & 13 & 2557 - 2597 &  A  & 41 & 1120 - 1132 &  B  & 13 &             &     &    \\
 1696 - 1706 &  B  & ~11 & 1130 - 1138 &  A  & ~9 &             &     &    & 1133 - 1139 & ABN & ~7 &             &     &    \\
 1707 - 1721 & ABN & ~15 & 1149 - 1161 &  A  & 13 &             &     &    & 1150 - 1160 &  B  & 11 &             &     &    \\
 1739 - 1750 &  B  & ~12 & 1177 - 1186 &  C  & 10 &             &     &    & 1167 - 1184 & ABN & 18 &             &     &    \\
 1761 - 1787 &  C  & ~27 & 1195 - 1212 &  B  & 18 &             &     &    & 1185 - 1219 &  A  & 35 &             &     &    \\
 1800 - 1812 &  B  & ~13 & 1221 - 1254 &  B  & 34 &             &     &    & 1221 - 1245 &  B  & 25 &             &     &    \\
 1813 - 1819 & ABN & ~~7 & 1255 - 1260 & ABN & ~6 &             &     &    & 1251 - 1277 &  C  & 27 &             &     &    \\
 1826 - 1835 &  C  & ~10 & 1281 - 1297 &  B  & 17 &             &     &    & 1297 - 1386 &  C  & 90 &             &     &    \\
 1847 - 2000 &  A  & 154 & 1305 - 1341 &  A  & 37 &             &     &    & 1395 - 1405 &  C  & 11 &             &     &    \\
 2004 - 2012 &  C  & ~~9 & 1351 - 1358 &  B  & ~8 &             &     &    & 1406 - 1415 & ABN & 10 &             &     &    \\
 2023 - 2035 & ABN & ~13 & 1364 - 1370 & ABN & ~7 &             &     &    & 1433 - 1437 &  C  & ~5 &             &     &    \\
 2043 - 2054 &  C  & ~12 & 1379 - 1387 & ABN & ~9 &             &     &    & 1444 - 1461 &  A  & 18 &             &     &    \\ 
 2058 - 2112 &  A  & ~55 & 1398 - 1410 &  A  & 13 &             &     &    & 1466 - 1476 &  B  & 11 &             &     &    \\
             &     &     &             &     &    &             &     &    &             &     &    &             &     &    \\
\hline
\end{tabular}
\end{table*}

\begin{table*}
\centering
\caption{Profile Widths and Emission Heights for 22 July, 2019 observations (see section~\ref{Emission height and geometry} for details).}
\label{tab:Profheight}
\begin{tabular}{cccccccccccc}
\hline
  Mode & Band & W$_{5\sigma}$ & W$_{10}$ & W$_{50}$ & W$_{sep}^{out}$ &  W$_{sep}^{in}$  & $\phi_{\circ}$ & $\phi_{t}^{5\sigma}$ & $\phi_{l}^{5\sigma}$ & $\Delta \phi$ & h$_{A/R}$\\
   &       &           &           &           &           &                 &           &                &        \\ 
   & (MHz) & ($\degr$) & ($\degr$) & ($\degr$) & ($\degr$) & ($\degr$) & ($\degr$) & ($\degr$)  & ($\degr$) & ($\degr$) & (km)\\ 
\hline
  A  & 300--330 & 30.9$\pm$0.2 & 29.0$\pm$0.2 & 23.5$\pm$0.2 &   19.6$\pm$0.2 & 6.4$\pm$0.2 & -8.7$\pm$0.3 & 15.1$\pm$0.3 & -15.8$\pm$0.3 & 0.7$\pm$0.3 & 329$\pm$141  \\
     & 330--360 & 35.8$\pm$0.2 & 27.9$\pm$0.2 & 23.2$\pm$0.2 &   19.6$\pm$0.2 & 5.6$\pm$0.2 & -8.7$\pm$0.2 & 17.2$\pm$0.2 & -18.5$\pm$0.2 & 1.3$\pm$0.2 & 611$\pm$94 \\
     & 379--409 & 34.2$\pm$0.2 & 27.1$\pm$0.2 & 22.4$\pm$0.2 &   17.6$\pm$0.2 & 4.4$\pm$0.2 & -8.6$\pm$0.2 & 16.0$\pm$0.3 & -18.2$\pm$0.3 & 2.2$\pm$0.3 & 1034$\pm$141 \\
     & 418--448 & 29.0$\pm$0.2 & 26.5$\pm$0.2 & 22.0$\pm$0.2 &   17.6$\pm$0.2 & 4.3$\pm$0.2 & -7.5$\pm$0.2 & 13.4$\pm$0.3 & -15.6$\pm$0.3 & 2.2$\pm$0.3 & 1034$\pm$141 \\
     & 455--485 & 31.8$\pm$0.2 & 25.9$\pm$0.2 & 21.7$\pm$0.2 &   17.6$\pm$0.2 & 4.4$\pm$0.2 & -7.8$\pm$0.2 & 15.4$\pm$0.2 & -16.4$\pm$0.2 & 1.0$\pm$0.2 & 470$\pm$94 \\
     &          &              &              &              &              &             &              &               &              &            \\
  B  & 300--330 & 32.3$\pm$0.2 & 32.2$\pm$0.2 & 24.2$\pm$0.2 &   20.4$\pm$0.2 & ---   & -9.1$\pm$0.6  & 14.0$\pm$0.6 & -18.3$\pm$0.6 & 4.3$\pm$0.6 & 2021$\pm$282  \\
     & 330--360 & 33.6$\pm$0.2 & 28.9$\pm$0.2 & 23.7$\pm$0.2 &   19.8$\pm$0.2 & ---   & -8.6$\pm$0.2   & 15.9$\pm$0.2 & -17.7$\pm$0.2 & 1.8$\pm$0.2 & 846$\pm$94 \\
     & 379--409 & 32.2$\pm$0.2 & 27.8$\pm$0.2 & 22.9$\pm$0.2 &   18.7$\pm$0.2 & ---   & -8.4$\pm$0.2  & 15.0$\pm$0.2 & -17.1$\pm$0.2 & 2.1$\pm$0.2 & 987$\pm$94 \\
     & 418--448 & 30.0$\pm$0.2 & 27.5$\pm$0.2 & 22.4$\pm$0.2 &   18.7$\pm$0.2 & ---   & -8.5$\pm$0.3  & 13.5$\pm$0.3 & -16.5$\pm$0.3 & 3.0$\pm$0.3 & 1410$\pm$141 \\
     & 455--485 & 30.1$\pm$0.2 & 26.8$\pm$0.2 & 22.3$\pm$0.2 &   18.5$\pm$0.2 & ---   & -8.5$\pm$0.2 & 14.1$\pm$0.2 & -16.0$\pm$0.2 & 1.9$\pm$0.2 & 893$\pm$94 \\
     &          &              &              &              &              &       &              &                &             &            \\
          
 ABN & 300--330 & 26.7$\pm$0.2 &   ---        & 24.0$\pm$0.2 & 20.1$\pm$0.2 &   ---  & 8.8$\pm$2.8  & 13.1$\pm$0.2 & -13.6$\pm$0.2 &  ---         & ---     \\
     & 330--360 & 32.3$\pm$0.2 & 29.5$\pm$0.2 & 23.7$\pm$0.2 & 19.9$\pm$0.2 &   ---  & 9.1$\pm$2.7  & 16.2$\pm$0.2 & -16.1$\pm$0.2 &  ---         & ---     \\
     & 379--409 & 30.1$\pm$0.2 & 27.8$\pm$0.2 & 22.9$\pm$0.2 & 18.5$\pm$0.2 &   ---  & 9.3$\pm$2.0  & 14.3$\pm$0.2 & -15.8$\pm$0.2 &  ---         & ---     \\
     & 418--448 & 28.4$\pm$0.2 &  ---         & 22.4$\pm$0.2 & 18.5$\pm$0.2 &   ---  & 10.0$\pm$2.0 & 13.1$\pm$0.2 & -15.3$\pm$0.2 &  ---         & ---     \\
     & 455--485 & 28.2$\pm$0.2 & 26.7$\pm$0.2 & 22.0$\pm$0.2 & 18.0$\pm$0.2 &   ---  & 9.5$\pm$2.0  & 13.0$\pm$0.2 & -15.3$\pm$0.2 &  ---         & ---      \\
     &          &              &              &              &              &        &              &               &              &          \\
  C  & 300--330 & 31.7$\pm$0.2 & 30.3$\pm$0.2 & 24.9$\pm$0.2 & 20.2$\pm$0.2 &   ---    & -9.9$\pm$0.2 & 15.3$\pm$0.2 & -16.4$\pm$0.2 & 1.1$\pm$0.2 & 517$\pm$94  \\
     & 330--360 & 34.4$\pm$0.2 & 29.3$\pm$0.2 & 24.3$\pm$0.2 & 19.8$\pm$0.2 &   ---    & -9.7$\pm$0.2 & 16.8$\pm$0.2 & -17.5$\pm$0.2 & 0.7$\pm$0.2 & 329$\pm$94 \\
     & 379--409 & 33.6$\pm$0.2 & 28.6$\pm$0.2 & 23.7$\pm$0.2 & 18.8$\pm$0.2 &   ---    & -8.9$\pm$0.2 & 16.4$\pm$0.2 & -17.2$\pm$0.2 & 0.8$\pm$0.2 & 376$\pm$94 \\
     & 418--448 & 30.1$\pm$0.2 & 28.1$\pm$0.2 & 23.2$\pm$0.2 & 18.5$\pm$0.2 &   ---    & -9.1$\pm$0.3 & 15.0$\pm$0.3 & -15.1$\pm$0.3 & --- & --- \\
     & 455--485 & 31.8$\pm$0.2 & 27.5$\pm$0.2 & 22.9$\pm$0.2 & 18.2$\pm$0.2 &   ---    & -8.9$\pm$0.2 & 16.0$\pm$0.3 & -15.8$\pm$0.3 &  --- & --- \\
     &          &              &              &              &              &             &              &                &              &         \\
 All & 300--330 & 33.4$\pm$0.2 & 29.5$\pm$0.2 & 24.0$\pm$0.2 & 19.8$\pm$0.2 &  ---  & -9.0$\pm$0.2  &  15.8$\pm$0.2 & -17.6$\pm$0.2 & 1.8$\pm$0.2 & 846$\pm$94  \\
     & 330--360 & 36.2$\pm$0.2 & 28.6$\pm$0.2 & 23.5$\pm$0.2 & 19.0$\pm$0.2 &  ---  & -8.7$\pm$0.2  & 17.6$\pm$0.2 & -18.6$\pm$0.2 & 1.0$\pm$0.2 & 470$\pm$94 \\
     & 379--409 & 35.3$\pm$0.2 & 27.6$\pm$0.2 & 22.8$\pm$0.2 & 18.5$\pm$0.2 &  ---  & -8.6$\pm$0.2  & 16.9$\pm$0.2 & -18.4$\pm$0.2 & 1.5$\pm$0.2 & 705$\pm$94 \\
     & 418--448 & 31.5$\pm$0.2 & 26.8$\pm$0.2 & 22.4$\pm$0.2 & 18.4$\pm$0.2 &  ---  & -8.6$\pm$0.2  & 14.8$\pm$0.2 & -16.8$\pm$0.2 & 2.0$\pm$0.2 & 940$\pm$94 \\
     & 455--485 & 33.3$\pm$0.2 & 26.5$\pm$0.2 & 22.0$\pm$0.2 & 17.7$\pm$0.2 &  ---  & -7.9$\pm$0.2  & 16.1$\pm$0.2 & -17.2$\pm$0.2 & 1.1$\pm$0.2 & 517$\pm$94 \\
     &          &              &              &              &              &             &              &                &              &          \\
     
\hline    
\end{tabular}
\end{table*}


\end{document}